\documentclass[prd,showpacs,showkeys,nofootinbib,floatfix,eqsecnum,
               fleqn,preprint,12pt,tightenlines]{revtex4-1} 

\usepackage{amsfonts,amsmath,amssymb}
\usepackage{bm}
\usepackage{verbatim}
\usepackage{graphicx}

\newcommand{\beq}{\begin{equation}}
\newcommand{\eeq}{\end{equation}}
\newcommand{\beqa}{\begin{eqnarray}}
\newcommand{\eeqa}{\end{eqnarray}}
\newcommand{\bsubeqs}{\begin{subequations}}
\newcommand{\esubeqs}{\end{subequations}}

\newcommand{\Chi}{X}   
\newcommand{\chempot}{\mu}   

\begin{document}

\begin{widetext}
%
%
\noindent Phys. Rev. D  \textbf{106}, 124015 (2022) 
\hfill    arXiv:2207.02826 
%
%
%
\newline\vspace*{3mm}
\end{widetext}

\title{Extension of unimodular gravity and the cosmological constant}

\author{\vspace*{5mm} F.R. Klinkhamer}
\email{frans.klinkhamer@kit.edu}
\affiliation{Institute for Theoretical Physics,
Karlsruhe Institute of Technology (KIT),\\ 76128 Karlsruhe,  Germany\\}

\begin{abstract}
\vspace*{2mm}\noindent
A new way is proposed to cancel
the cosmological constant.
The proposal involves the metric determinant
acting as a type of self-adjusting $q$-field
without need of a fine-tuned chemical potential.
Since the determinant of the metric now plays a role in the physics,
the allowed coordinate transformations are restricted to
those with unit Jacobian. This approach 
to the cosmological constant problem is, therefore, similar
to the unimodular-gravity approach of the previous literature.
The resulting cosmology has been studied and
the obtained results show the natural
cancellation of an initial cosmological constant
if quantum-dissipative effects are included.
\end{abstract}

\maketitle

\section{Introduction}
\label{sec:Introduction}

The cosmological constant problem,  
at the interface of gravitation
and elementary particle physics,
is certainly one of the most important
problems of modern physics~\cite{Weinberg1989,Carroll2001}.
In a nutshell, the main cosmological constant problem is as follows
(throughout, we use natural units with $c=1$  and $\hbar=1$).
The electroweak standard model of elementary particles
involves a vacuum energy density $\epsilon_{V}^\text{\,(EWSM)}$
of the order of $(100\;\text{GeV})^4 \sim 10^{44}\;\text{eV}^4$.
Moreover, this energy density can be expected to
vary as the temperature $T$ of the Universe drops.
How, then, can the Universe end up with a vacuum energy density
$\Lambda^\text{(observed)}$ of order $10^{-11}\;\text{eV}^4$?
There are 55 orders of magnitude to explain. 
See, in particular, Ref.~\cite{Carroll2001}  
for further discussion of the astronomical observations.

We remark that the cosmological constant problem is about 
canceling all different contributions to the 
vacuum energy density 
appearing over the whole history of the Universe
and not about canceling just one number.
For this reason, some form of adjustment mechanism
seems to be called for.
A particular adjustment mechanism has been proposed
that is inspired by condensed matter physics, 
where a special type of vacuum variable $q$ provides
for the natural cancellation of any previously available
vacuum energy density~\cite{KlinkhamerVolovik2008a,KlinkhamerVolovik2008b}.

The 4-form realization of $q$-theory suggests the existence
of a chemical potential $\chempot$
which leads to the cancellation of the
gravitating vacuum energy density in equilibrium.
However, the dynamics towards the equilibrium remains a problem  
since the relaxation of $\chempot$ to its equilibrium value $\chempot_{0}$
in the Minkowski vacuum was not demonstrated in the original
papers~\cite{KlinkhamerVolovik2008a,KlinkhamerVolovik2008b}.

Here, we consider another version of $q$-theory,
where the role of the dynamical vacuum variable $q$
is played by the tetrad determinant~\cite{KlinkhamerVolovik2019}.
The chemical potential in this case may arise, for example,
from a model of the vacuum as a spacetime crystal,
where the number of lattice points is conserved, which
gives rise to a chemical potential $\chempot$.

The present paper essentially consists of two parts,   
where the second part (Secs.~\ref{sec:Cosmology-Basic-model} 
and \ref{sec:Cosmology-Quantum-dissipative-effects}) presents
the main cosmological results from an assumed action \eqref{eq:action-S}.
This assumed action, with only the fields of  general relativity and the 
standard model of elementary particles, has a single
nonstandard term involving 
the square root of the negative metric determinant
(or, equivalently, the tetrad determinant).
The first part 
(Secs.~\ref{sec:Crystal}--\ref{sec:Comparison-with-cond-mat})
gives a condensed-matter-inspired motivation 
for the assumed action \eqref{eq:action-S}, 
but this action may very well have another origin.  

The specific content of the first part,
which can be skipped in a first reading, is as follows.   
In Sec.~\ref{sec:Crystal}, we present a physical motivation
for having a chemical potential associated with the metric determinant.
In Sec.~\ref{sec:Cancellation-of-CC}, we then show that
the metric determinant can, in principle, cancel an initial
cosmological constant, but the chemical potential needs
to be fine-tuned.
In Sec.~\ref{sec:Metric-determinant-as-dynamic-variable},
we avoid this fine-tuning of the chemical potential
by introducing a nonstandard coupling of the
metric determinant to matter,
provided the allowed coordinate transformations are restricted to
those with unit Jacobian.
In Sec.~\ref{sec:Comparison-with-cond-mat}, we compare
the new metric-determinant cancellation mechanism with
what happens in condensed matter physics,  which was
the inspiration of our previous work on $q$-theory
(for a brief review,
see App.~A in Ref.~\cite{KlinkhamerVolovik2022-BBasTopQPT}).

The specific content of the second part,
which can essentially be read without knowledge of the first part, 
is as follows. 
In Sec.~\ref{sec:Cosmology-Basic-model}, we present a basic
model for cosmology with the
metric determinant as a dynamic variable.
We have both analytic and numeric results, but
the cosmological constant cannot be cancelled in general.
For that cancellation, we may need to appeal to nonreversible
effects such as dissipation.
In Sec.~\ref{sec:Cosmology-Quantum-dissipative-effects},
we consider a phenomenological model of
cosmology with quantum-dissipative effects included.
The main result is that it appears possible
to cancel a cosmological constant for initial boundary conditions 
within a finite domain (attractor behavior).
The three appendices give further results. 

In Sec.~\ref{sec:Final-remarks}, we give some concluding 
remarks on both parts.

\section{Spacetime crystal: Conservation of lattice points}
\label{sec:Crystal}

As explained in Sec.~\ref{sec:Introduction}, it is possible
to skip ahead to Sec.~\ref{sec:Cosmology-Basic-model} in a first reading.
The present section sets out to explore a potential
condensed-matter-type origin of the action used later for cosmology.

For a $(3+1)$-dimensional vacuum crystal
with elasticity tetrads~\cite{NissinenVolovik2019,Nissinen2020}
\begin{equation}
E^{\,a}_{\alpha}(x) = \partial_{\alpha} \, X^{a}(x)  \,,
\label{E}
\end{equation}
the density $n$ of lattice points is determined 
by the volume of the Brillouin zone,  
\begin{equation}
 n(x) =\frac{1}{(2\pi)^{4}}\,  e^{\alpha\beta\gamma\delta}\,
 E^{\,0}_{\alpha}(x)\, E^{\,1}_{\beta}(x)\, E^{\,2}_{\gamma}(x)\, E^{\,3}_{\delta}(x)\,.
\label{n}
\end{equation}
Neglecting factors of $2\pi$, the quantity $n(x)$ 
equals the tetrad determinant $E(x)$ with the dimension of inverse length
to the fourth power
(the phase fields $X^{a}$ are dimensionless).

As suggested in Sec.~VII of Ref.~\cite{NissinenVolovik2019},
it is, in principle, possible that gravity emerges from
a vacuum crystal with elasticity tetrads \eqref{E}.
There would then be an effective metric $g_{\alpha\beta}$ built
from the elasticity tetrads,
$g_{\alpha\beta} \propto \eta_{a b}\,E^{\,a}_{\alpha}\,E^{\,b}_{\beta}\,$.
In that case, we can identify the tetrad determinant
with the square root of minus the metric determinant, $E\propto\sqrt{-g}$.
If the vacuum crystal has a
fundamental length scale $\ell \equiv 1/M$, we would have
\beq
\label{eq:n-E-sqrtminusg}
M^{-4}\,n(x)\equiv M^{-4}\,E(x)=\sqrt{-g(x)}\,,
\eeq
which relates the lattice-point density $n$
to the emergent dimensionless metric $g_{\alpha\beta}$ that
enters the Einstein--Hilbert action for gravity.

The total number of lattice points is given by
\begin{equation}
N=\int d^{4}x  \, n(x)\,,
\label{N}
\end{equation}
and it is natural to assume that this number is conserved.
Then there is a Lagrange multiplier in the action,  
\begin{equation}
S_{N}=-\chempot \, N=- \chempot \int d^{4}x  \; n(x)\,,
\label{SN}
\end{equation}
where $\chempot$ is the corresponding chemical potential. 
This chemical potential $\chempot$ is dimensionless, 
which may be of direct relevance
for a recent proposal to replace the big bang singularity 
by a quantum phase transition~\cite{KlinkhamerVolovik2022-BBasTopQPT}.

Instead of a spacetime crystal
with a conserved number of lattice points,
one might consider the conventional vacuum,
but now with conservation of the 4-volume.
This would give the same action (\ref{SN})
with $n$ from \eqref{eq:n-E-sqrtminusg},
which is really the only input needed 
for the following.

\section{Cancellation of the cosmological constant}
\label{sec:Cancellation-of-CC}

The total action is
\begin{equation}
S =  S_{G} + S_{M} + S_{N}\,,
\label{TotalAction}
\end{equation}
with the action $S_{N}$ from \eqref{SN},
the gravitational Einstein--Hilbert action $S_{G}$ 
containing the Ricci curvature scalar, 
and the action $S_{M}$ for the matter fields.
The action (\ref{TotalAction}) is fully diffeomorphism invariant.

To study the Minkowski vacuum, we can neglect the gradient terms
in the matter action and also the curvature term.
The matter term then depends only on a potential $\epsilon\,$:%
\begin{equation}
S_{M} =\int d^{4}x \, M^{-4}\,n \, \epsilon(\Phi)\,,
\label{MatterAction}
\end{equation}
where $\Phi$ is a generic scalar field, considered here 
to be without gradients, i.e., constant over the spacetime manifold.
The equilibrium vacuum state is obtained by variation
of this last action over $\Phi$:%
\begin{equation}
\left.\frac{d \epsilon(\Phi)}{d\Phi}\right|_{\Phi=\Phi_{0}}=0\,.
\label{Equilibrium}
\end{equation}
The equilibrium value $\Phi_{0}$ gives the vacuum energy
contribution  $\epsilon(\Phi_{0})$ to the effective cosmological constant.
It is nonzero if there is no artificial fine-tuning.

The total contribution to the vacuum energy density $\rho_\text{vac}$
that enters the Einstein gravitational field
equation comes from $S_{M}+S_{N}$ by variation of $M^{-4}\,n$,  
\begin{equation}
\rho_\text{vac}= \epsilon(\Phi_{0}) - \chempot \, M^{4}\,.
\label{VacEnergyLambda}
\end{equation}
From the Einstein equation applied to the state with zero curvature $R=0$
and zero temperature $T=0$,
we obtain the equilibrium value $\chempot=\chempot_{0}$ 
of the Minkowski vacuum,  
 \begin{equation}
\rho_\text{vac}= \epsilon(\Phi_{0}) - \chempot_{0}\,M^{4} =0 \,.
\label{VacEnergyNullification}
\end{equation}
The vacuum energy density of the matter field $\epsilon(\Phi_{0})$  is naturally
cancelled by the chemical potential $\chempot=\chempot_{0}$.
This is similar to the 4-form $q$-theory~\cite{KlinkhamerVolovik2008a},
where  the equations
$\epsilon(q)  -\chempot \, q =0$ and $d\epsilon/dq -\chempot =0$ determine
both  $q_{0}$ and $\chempot_{0}$.

The main problem, now, is in the dynamics:  
how to describe the dynamical
relaxation of the parameter $\chempot$ to its equilibrium
value $\chempot_{0}$.

\section{Metric determinant as a dynamic variable}
\label{sec:Metric-determinant-as-dynamic-variable}

Probably the best way to deal with the chemical-potential
fine-tuning problem
is to introduce the dependence of the matter energy density on $n$,
i.e, to have a potential $\epsilon=\epsilon(\Phi,\,n)$.
In this case, the action
is invariant only under those coordinate transformations that have
a Jacobian equal to unity,
\beq
\label{eq:Jacobian-unity}
\det\Big(\partial x'^{\,\alpha}/\partial x^{\beta}\Big) =1\,.
\eeq
These restricted coordinate transformations also appear
in the unimodular-gravity approach to the cosmological constant 
problem~\cite{Einstein1919,vanderBij-etal1982,Zee1983,BuchmuellerDragon1988,%
HenneauxTeitelboim1989} (a brief review is given in
Sec.~VII of Ref.~\cite{Weinberg1989}).

The Minkowski vacuum may then have a continuous set of $\chempot$ values, which
determine the equilibrium values $n_{0}(\chempot)$ of the metric determinant
in equilibrium.  In other words, there are many different quantum vacua
in flat Minkowski spacetime and they are parametrized   %
by the values $\chempot$ of the chemical potential.
(Recall that the Minkowski vacuum in the original
$q$-theory~\cite{KlinkhamerVolovik2008a} has a single value
$\chempot_{0}$ for the chemical potential.)

The Einstein gravitational field equation (to be given explicitly
in Sec.~\ref{subsec:Cosmology-Basic-model-Action-Ansaetze})
now contains the following vacuum energy density:
\begin{equation}
\rho_\text{vac}(\Phi,n) =M^{4}\,
\frac{d}{dn}\Big[n\,M^{-4}\,\epsilon(\Phi,\,n)-\chempot \, n\Big]=
\epsilon(\Phi,\,n) 
+ n\, \frac{d \epsilon(\Phi,\,n)}{dn} - \chempot\, M^{4}\,.
\label{VacEnergy}
\end{equation}
The equilibrium vacua are obtained by variation of the action
over both $\Phi$ and $n$,  
\bsubeqs\label{eq:Equilibrium1and2}
\beqa
\frac{d \epsilon(\Phi,\,n)}{d\Phi}&=&0 \,,
\label{Equilibrium1}
\\[2mm]
\rho_\text{vac}(\Phi,n)&=&0\,.
\label{Equilibrium2}
\eeqa
\esubeqs
These equations determine the equilibrium values
of the variables $\Phi$ and $n$ as functions of $\chempot$,
that is, having $\Phi=\Phi(\chempot)$ and  $n=n(\chempot)$
over a finite range of $\chempot$.
This range does not necessarily include $\chempot=0$,
as can be seen from the example below.
Hence, it is necessary to introduce $\chempot$
and we cannot just forget about it.

The simplest example is  
\bsubeqs\label{eq:example-epsilon-rhoV}
\beqa
\epsilon(\Phi,\,n) &=&
\widetilde{\epsilon}(\Phi)\left(1+ \frac{n}{n_\text{scale}} \right)\,,
\label{Expansion3}
\eeqa
with a fixed positive density $n_\text{scale}$
(alternatively written as $M^{4}$). The corresponding 
gravitating vacuum energy density  
from \eqref{VacEnergy} reads 
\beqa
\rho_\text{vac}(\Phi,n) &=&
\widetilde{\epsilon}(\Phi)
\left(1+ 2 \,\frac{n}{n_\text{scale}} \right)- \chempot \,M^{4}\,.
\label{Expansion4}
\eeqa
\esubeqs
Now, condition (\ref{Equilibrium1}) gives the equilibrium value $\Phi_{0}$
of the matter field $\Phi$ and condition (\ref{Equilibrium2})
gives the equilibrium value $n_{0}=n_{0}(\chempot)$
of the metric determinant $n \propto \sqrt{-g}$,  
\begin{eqnarray}
n_{0}(\chempot)=n_\text{scale} \;
\frac{\chempot\,M^{4}
-\widetilde{\epsilon}(\Phi_{0})}{2\,\widetilde{\epsilon}(\Phi_{0})} \,.
\label{Expansion5}
\end{eqnarray}
Assume, for definiteness, that $\widetilde{\epsilon}(\Phi_{0})>0$.
Then, self-sustained  Minkowski vacua 
with $n_{0}>0$     
exist only at $\chempot\,M^{4} > \widetilde{\epsilon}(\Phi_{0})$, 
which does not allow for $\chempot=0$ as mentioned above.
[Incidentally,      
the condition $\chempot\,M^{4} > \widetilde{\epsilon}(\Phi_{0})$
can be relaxed by modifying the $\epsilon(\Phi,\,n)$ 
\textit{Ansatz} \eqref{Expansion3};
a related example will be presented in the last
paragraph of App.~\ref{app:Analytic-solution-wchi-general}.]

This approach with a dynamically-fixed metric determinant
is an extension of the unimodular-gravity
approach~\cite{Einstein1919,vanderBij-etal1982,Zee1983,BuchmuellerDragon1988,%
HenneauxTeitelboim1989},
where typically the metric determinant is eliminated as a  
dynamical variable; see, e.g., the second and third paragraphs
of Sec.~VII in Ref.~\cite{Weinberg1989}. 
(Some related ideas 
on a dynamical measure of integration in the action,
generalizing $\sqrt{-g}$,
appear in Ref.~\cite{BensityGuendelman-etal2020}  
and references therein.) 

The simple model \eqref{eq:example-epsilon-rhoV}
can be used for calculations of the dynamics of the cosmological constant, 
since, contrary to the original $q$-theory, the relaxation to the Minkowski
vacuum  does not require the fine-tuning of $\chempot$ 
to the value $\chempot_{0}$. 
Instead of the problematic relaxation of $\chempot$, 
there is the relaxation of $n$ 
at a given value $\chempot\,M^{4} > \widetilde{\epsilon}(\Phi_{0}) > 0$ 
for the example considered.  

We can also study the relaxation after a cosmological phase transition, 
at which the $\Phi_{0}$ value may  change.  
Different from the original $q$-theory approach, 
this does not require a change in the chemical potential. 
After the transition, the quantity $n_{0}$ will be adjusted 
to a new equilibrium state, while $\chempot$ remains fixed.  
Moreover, we can expect a phase transition 
between the self-sustained Minkowski vacuum 
with $\widetilde{\epsilon}(\Phi_{0})< \chempot\,M^{4}$ 
and the state with $\widetilde{\epsilon}(\Phi_{0})> \chempot\,M^{4}$, 
which can be expanding or contracting.

\section{Comparison with condensed matter physics}
\label{sec:Comparison-with-cond-mat}

In condensed matter physics with conservation of particle number,
there is the thermodynamic equilibrium equation,%
\bsubeqs\label{eq:equilibriumequation-GibbsDuhem}
\beqa
\label{eq:equilibriumequation}
\frac{d \epsilon}{d n} =\chempot\,,
\eeqa
and the Gibbs--Duhem relation at temperature $T=0$,%
\beqa
 \rho_\text{vac}=\epsilon(n) -\chempot \, n =-P\,.
\label{GibbsDuhem}
\eeqa
\esubeqs
The combination of both relations determines
$\chempot$ and $n$ as a function of the pressure $P$.

If there is no external pressure, $P_\text{external}=0$, one obtains
the nullification of the effective cosmological constant,%
\begin{equation}
\rho_\text{vac}=  \epsilon(n)- \chempot \, n=-P=-P_\text{external}=0\,.
\label{Nullification}
\end{equation}
Here,
the nullification of the vacuum energy is provided
by the absence of an external environment.
But only self-sustained systems can exist 
in the absence of external pressure.

For the example \eqref{eq:example-epsilon-rhoV}, self-sustained vacua exist
if $\chempot\,M^{4} > \epsilon(\Phi_{0})$.
The vacua with $\chempot\,M^{4} <\epsilon(\Phi_{0})$ are not self-sustained,
so that
the cosmological constant (the analog of the external pressure) is nonzero
and leads to expansion or contraction of these vacua
(a de-Sitter-type universe).

\section{Cosmology: Basic model}
\label{sec:Cosmology-Basic-model}

\subsection{Action and Ans\"{a}tze}  
\label{subsec:Cosmology-Basic-model-Action-Ansaetze}

We will now investigate the application to cosmology
of the theory as discussed in the previous sections.
These previous sections
can, however, be skipped in a first reading,  
as the present section and the next are self-contained.  
Here, we aim to establish
the asymptotic vanishing of the total gravitating vacuum energy density.
For that, we simplify the theory to the
bare minimum: we remove the scalar $\Phi$ field (which is 
not really needed for the cosmological-constant cancellation)
and add a standard real scalar $\Chi$
(needed to get the appropriate expansion of the
Friedmann--Robertson--Walker-type model).

The postulated action is given by
\bsubeqs\label{eq:action-S}
\beqa
S &=& S_{G}+ S_{M} + S_{\Lambda-\text{plus}} + S_{N}\,,
\\[1.000mm]
\label{eq:action-G}
S_{G}&=& \int d^{4}x  \,\sqrt{-g}\;\frac{R}{16\pi G_{N}}\,,
\\[1.000mm]
\label{eq:action-SChi}
S_{M}&=& \int d^{4}x  \,\sqrt{-g}\;
\left[  \frac{1}{2}\,g^{\alpha\beta}\, \partial_{\alpha}\Chi\,\partial_{\beta}\Chi
+  \overline{\epsilon}(\Chi) \right]\,,
\\[1.000mm]
\label{eq:action-Lambdaplus}
S_{\Lambda-\text{plus}}&=& \int d^{4}x  \,\sqrt{-g}\;
  \epsilon(\Lambda,\,n) \,,
\\[1.000mm]
\label{eq:action-SN}
S_{N} &=& - \chempot\, \int d^{4}x  \; n(x)\,,
\\[1.000mm]
\label{eq:action-n-def}
n(x)&=& \sqrt{-g(x)} \;  M^{4} \,,
\eeqa
\esubeqs
where $g(x)$ is the determinant of the metric $g_{\alpha\beta}(x)$
with Lorentzian signature and $1/M$ is 
a fundamental length scale of the underlying theory.  
In \eqref{eq:action-SChi} and
\eqref{eq:action-Lambdaplus},
we simply take
\bsubeqs\label{eq:epsilonbar-epsilon-Ansaetze}
\beqa
\label{eq:epsilonbar-Ansatz}
\overline{\epsilon}(\Chi) &=&
\frac{1}{2}\,g_2\,M^{2}\,\Chi^{2}  \,,
\\[2mm]
\label{eq:epsilon-Ansatz}
\epsilon(\Lambda,\,n) &=& \Lambda  + \zeta\,n \,,
\eeqa
\esubeqs
with real parameters  $g_2\geq 0$ and $\zeta>0$.
We emphasize that, strictly speaking, 
the only new input is the single term $n \propto \sqrt{-g}$ 
in the potential \eqref{eq:epsilon-Ansatz},
consistent with having coordinate invariance restricted by
\eqref{eq:Jacobian-unity}. A possible 
condensed-matter-type origin of the action \eqref{eq:action-S}
has been discussed in 
Secs.~\ref{sec:Crystal}--\ref{sec:Comparison-with-cond-mat}, 
but this action can also have an entirely different origin.

In the resulting gravitational field equation,
\beq
\label{eq:Einstein-eq}
\frac{1}{8\pi G_{N}}
\left( R_{\alpha\beta}-\frac{1}{2}\,R\,g_{\alpha\beta}\right)=
\rho_\text{vac}\, g_{\alpha\beta}+T^{M}_{\alpha\beta}\,,
\eeq
we have 
\bsubeqs\label{eq:rhovac-mu-Ansaetze}
\beqa
\rho_\text{vac} &=&
\Lambda  + 2\,\zeta\,n -\chempot\,M^{4} \,,
\\[2mm]
\Lambda&=&  \lambda\,M^{4}\,,
\eeqa
\esubeqs
where the chemical potential $\chempot \ne 0$ traces back to
the action term \eqref{eq:action-SN}
and $n$  has been defined by \eqref{eq:action-n-def}.  
Taking the covariant divergence of \eqref{eq:Einstein-eq}
and using the contracted Bianchi identities, 
we obtain 
the following combined energy-momentum conservation relation: 
\beq
\label{eq:combined-energy-momentumconservation}
\Big( \rho_\text{vac}\, g_{\alpha\beta}+T^{M}_{\alpha\beta}\Big)^{;\,\beta}
=0\,,
\eeq
where the semicolon stands for a covariant partial derivative
(the colon stands for a standard partial derivative).    
If the matter component is separately conserved,
$\left( T^{M}_{\alpha\beta}\right)^{;\,\beta} =0$,
then equally so for the vacuum component, so that
$\rho_\text{vac}^{\hspace*{4mm},\,\beta} =0$.

With diffeomorphisms restricted to those of unit
Jacobian, the appropriate spatially-flat
Robertson--Walker (RW) metric has been given 
in Ref.~\cite{AlvarezFaedo2007} (see also Ref.~\cite{Zee1983}):
\beqa\label{eq:extRW-ds2}
\hspace*{-0mm}
ds^{2}
&=&
g_{\alpha\beta}(x)\, dx^{\alpha}\,dx^{\beta}
=
- \widetilde{A}(t)\;d t^{2}
+ \widetilde{R}^{\,2}(t)\;\delta_{i j}\,dx^{i}\,dx^{j}\,,
\eeqa
where $t$ is the cosmic time coordinate from $x^{0}=c\,t=t$
and $\widetilde{A}(t)>0$ 
is an additional \textit{Ansatz} function.
The spatial indices $i$, $j$ in \eqref{eq:extRW-ds2}
run over $\{1,\, 2,\, 3 \}$
and $\widetilde{R}(t)$ is the cosmic scale factor
[the tilde marks the difference with the Ricci scalar
appearing in \eqref{eq:action-G}].
For $\widetilde{A}(t)=\text{const}>0$, we recover the standard
spatially-flat RW metric.
We remark that the extended RW metric \eqref{eq:extRW-ds2}    
gives the vacuum variable
\beq\label{eq:n-for-ext-RW-metric}
n \propto \sqrt{-g} =(\widetilde{A}\,)^{1/2}\:|\widetilde{R}\,|^{3}\,,
\eeq
with proportionality constant $M^4$ according to
\eqref{eq:action-n-def}.

If the scalar field $\Chi$ is spatially homogeneous in the
cosmological spacetime \eqref{eq:extRW-ds2},
$\Chi=\Chi(t)$, then its energy-momentum tensor corresponds to that of
a perfect fluid with the following energy density 
and pressure~\cite{Mukhanov2005}:%
\bsubeqs\label{eq:rho-P-from-scalar}
\beqa
\rho_{\Chi}(t) &=&
\frac{1}{2}\,\frac{1}{a(t)}\,\left(\frac{d\Chi(t)}{d t}\right)^{2}
+ \frac{1}{2}\,g_2\,M^{2}\,\Big(\Chi(t)\Big)^{2}\,,  
\\[2mm]
P_{\Chi}(t) &=&
\frac{1}{2}\,\frac{1}{a(t)}\,\left(\frac{d\Chi(t)}{d t}\right)^{2} 
- \frac{1}{2}\,g_2\,M^{2}\,\Big(\Chi(t)\Big)^{2}\,.
\eeqa
If the scalar field $\Chi$ is, moreover, rapidly oscillating,
$\Chi(t)=\Chi_{0}\,\cos(\omega\,t)$, then
the time averages of the energy density and the pressure
give the following matter equation-of-state parameter:
\beqa
w_{M} &=&
\frac{\langle P_{\Chi}\rangle}{\langle \rho_{\Chi}  \rangle}
=
\frac{\omega^{2}/a - g_2\,M^{2}}{\omega^{2}/a + g_2\,M^{2}}\,,
\eeqa
\esubeqs
where the cosmological time scale relevant to $a(t)$
is assumed to be much larger than $1/\omega$ or $1/M$.
Obviously, $g_2=0$ gives $w_{M} =1$ and
a value $w_{M} =1/3$ follows from $\omega^{2}/a= 2\,g_2\,M^{2}$.

In the following, we will work with this perfect fluid
instead of the original scalar $\Chi$ field
and take $w_{M} =1/3$, which can be interpreted
as a gas of ultrarelativistic particles.

\subsection{Dimensionless ordinary differential equations}  
\label{subsec:Cosmology-Basic-model-ODEs}

Henceforth, we set
\beq
\label{eq:EPlanck-M-unity}
M = E_\text{Planck} \equiv 1/\sqrt{G_{N}}\,,
\eeq
and introduce the following dimensionless quantities
(the chemical potential $\chempot$ is already dimensionless):%
\bsubeqs\label{eq:dimensionless-variables}
\begin{align}
t &\to \tau\,,
\hspace*{-10mm}
&\rho_{\Chi}(t) &\to r_{\chi}(\tau)\,,
\hspace*{-10mm}
&\widetilde{A}(t) &\to a(\tau)\,,
\\[2mm]
\Chi(t) &\to \chi(\tau)\,,
\hspace*{-10mm}
&P_{\Chi}(t) &\to p_{\chi}(\tau)\,,
\hspace*{-10mm}
&\widetilde{R}(t) &\to r(\tau)\,,
\\[2mm]
n(t) &\to n(\tau)\,,
\hspace*{-10mm}
&\Lambda &\to \lambda\,,
\hspace*{-10mm}
&
\end{align}
\esubeqs
where $n(\tau)$ is dimensionless and equal to
$\sqrt{-g(\tau)}=\sqrt{a(\tau)}\,|r(\tau)|^{3}$.

From the field equations of the action \eqref{eq:action-S}
and using the homogeneous perfect fluid from the $\chi$ scalar,
we obtain the following
dimensionless ordinary differential equations (ODEs):
\bsubeqs\label{eq:dimensionless-ODEs}
\beqa
\label{eq:dimensionless-ODE-rchidoteq}
\hspace*{0mm}
&&\dot{r}_{\chi}
+  3\,(1+w_{M})\, \left(\frac{\dot{r}}{r}\right) \,r_{\chi}
 = 0 \,,
\\[2mm]
\label{eq:dimensionless-ODE-1stF}
\hspace*{0mm}
&&3\,\left( \frac{\dot{r}}{r} \right)^{2}
 =
8\,\pi \,a\, \Big(r_{\chi}+r_\text{vac} \Big)\,,
\\[2mm]
\label{eq:dimensionless-ODE-2ndF}
\hspace*{0mm}
&&\frac{2\,\ddot{r}}{r}
+\left( \frac{\dot{r}}{r} \right)^{2}
-\left( \frac{\dot{a}}{a} \right)\,\left( \frac{\dot{r}}{r} \right)
 =
- 8\,\pi \,a\,  \Big( w_{M}\,r_{\chi}-  r_\text{vac} \Big) \,,
\\[2mm]
\label{eq:dimensionless-ODE-rvac}
\hspace*{0mm}
&&r_\text{vac}  =
\lambda   + 2\,\zeta\,\sqrt{a}\,|r|^{3}  -\chempot \,,
\eeqa
\esubeqs
where the overdot stands for
differentiation with respect to $\tau$.
These ODEs have three real parameters:  
the matter equation-of-state parameter $w_{M} > -1$
and two parameters
entering the vacuum energy density $r_\text{vac}$,
namely $\zeta>0$ and $\mu \ne  0$. 
Incidentally, the function $a(\tau)$ has been assumed to be positive,
so that there is no difficulty in taking its square root.

It can be shown that the ODEs \eqref{eq:dimensionless-ODEs}
give an equation for the constancy of the vacuum energy density,
\beq
\label{eq:dimensionless-ODE-rvacdot-zero}
\dot{r}_\text{vac}=0\,.
\eeq
This equation corresponds to
the energy-conservation equation of a homogeneous perfect
fluid with equation-of-state parameter $w_\text{vac}=-1$,
compared with \eqref{eq:dimensionless-ODE-rchidoteq}
for the matter component.
In fact, \eqref{eq:dimensionless-ODE-rvacdot-zero} traces back
to  \eqref{eq:combined-energy-momentumconservation}
for matter with $\left( T^{M}_{\alpha\beta}\right)^{;\,\beta} =0$,
so that $\rho_\text{vac}^{\hspace*{4mm},\,\beta} =0$.
In Sec.~\ref{sec:Cosmology-Quantum-dissipative-effects},
we will introduce a vacuum-matter energy exchange,
but here we do without and keep \eqref{eq:dimensionless-ODE-rvacdot-zero}.

\subsection{Analytic Friedmann-type solution for $\mathbf{w_{M}=1/3}$}
\label{subsec:Analytic-solution-wchi-onethird}

We now present an exact solution of the ODEs \eqref{eq:dimensionless-ODEs}
for $w_{M}=1/3$ (an exact solution for general $w_{M}>-1$
is given in App.~\ref{app:Analytic-solution-wchi-general}).
We take the following \textit{Ansatz} functions for $\tau > 0$:  
\bsubeqs\label{eq:functions-analytic-sol}
\beqa
a(\tau) &=& \alpha\;\tau^{-2\,p} \,,
\\[2mm]
r(\tau) &=& \alpha^{-1/6}\;\widehat{r}\; \tau^{p/3} \,,
\\[2mm]
r_{\chi}(\tau) &=& \alpha^{-1}\;\widehat{\chi}\;\tau^{-m}\,,
\eeqa
\esubeqs
with positive parameters $\alpha$, $p$, $\widehat{r}$, $\widehat{\chi}$,
and $m$. The vanishing of $r_\text{vac}$ 
from \eqref{eq:dimensionless-ODE-rvac} gives immediately
\beq
\label{eq:rhat-analytic-sol}
\widehat{r}_\text{\,sol} =
\sqrt[3]{\frac{\,1}{2\,\zeta}\;\big(\chempot- \lambda\big)}\,,
\eeq
where, for a given value $\chempot>0$, the dimensionless cosmological constant
$\lambda$  must obey the following condition:
\beq
\label{eq:lambda-condition-analytic-sol}
 \lambda < \chempot\,,
\eeq
so that $\lambda$ can also be negative. Here, and in the following,
we have assumed a positive $\chempot$, but similar results
are obtained for a negative $\chempot$.

For the  \textit{Ansatz} functions \eqref{eq:functions-analytic-sol},
the dimensionless Ricci and Kretschmann curvature scalars read%
\bsubeqs\label{eq:R-K-analytic-sol}
\beqa
\label{eq:R-analytic-sol}
\mathcal{R}&=&
\frac{2}{3}\,
p\,\big(  5\,p -3 \big) \,\frac{1}{\alpha}\,{\tau}^{-2\,(1- p)}\,,
\\[2mm]
\label{eq:K-analytic-sol}
\mathcal{K}&=&
\frac{4}{27}\,p^{2}\,\big( 9 - 24\,p + 17\,p^{2} \big) \,
\frac{1}{\alpha^2}\,{\tau}^{-4\,(1- p)}\,.
\eeqa
\esubeqs
Restricting the power $p$ of the  \textit{Ansatz} functions to the range
\beq
\label{eq:p-bounds-analytic-sol}
0< p < 1\,,
\eeq
we look for an expanding ($p>0$) Friedmann-type
universe
approaching Minkowski spacetime (different from a
de-Sitter spacetime at $p=1$; cf. Eq.~(52) in Ref.~\cite{Bamba-etal2016}).
Some details of the analytic de-Sitter-type solution 
with $p=1$ and $\widehat{\chi}=0$ are given in 
App.~\ref{app:Analytic-solution-deS}.

With the \textit{Ansatz} functions \eqref{eq:functions-analytic-sol},
the three ODEs from \eqref{eq:dimensionless-ODEs}
reduce to the following expressions:
\bsubeqs\label{eq:ODES-functions-analytic-sol}
\beqa
\label{eq:ODES-rchieq-functions-analytic-sol}
0 &=&
\frac{1}{\alpha}\,
\left(  4\,p/3  - m \right) \,\widehat{\chi}\,{\tau}^{-1 - m} \,,
\\[2mm]
\label{eq:ODES-1stFeq-functions-analytic-sol}
0 &=&
\frac{p^{2}}{3\,{\tau}^{2}} - 8\,\pi \,\widehat{\chi}\,{\tau}^{-m - 2\,p}\,,
\\[2mm]
\label{eq:ODES-2ndFeq-functions-analytic-sol}
0 &=&
\frac{p^{2}}{{\tau}^{2}} -\frac{2\,p}{3\,{\tau}^{2}} +
\frac{8}{3}\,\pi \,\widehat{\chi}\,{\tau}^{-m - 2\,p}\,.
\eeqa
\esubeqs
These equations have an exact solution,  
\bsubeqs\label{eq:constants-analytic-sol}
\beqa
p_\text{\,sol} &=& 3/5\,,
\\[2mm]
m_\text{\,sol} &=& 4/5\,,
\\[2mm]
\widehat{\chi}_\text{\,sol} &=& \frac{3}{200\,\pi } \approx 0.00477465 \,,
\eeqa
\esubeqs
for arbitrary $\alpha>0$.
Technically speaking, the solution exists because of an interplay
between the two metric \textit{Ansatz} functions
and the matter energy density. The evaluated matter equation
\eqref{eq:ODES-rchieq-functions-analytic-sol}
fixes the power $m$ to be equal to $4\,p/3$.
The evaluated first Friedmann equation
\eqref{eq:ODES-1stFeq-functions-analytic-sol} then fixes  
$p=3/5$ and $m=4/5$, so that both terms
in the equation get
the same temporal dependence $\tau^{-2}$ and
cancel each other for an appropriate value of the
constant $\widehat{\chi}$.
It turns out that the evaluated second Friedmann equation
\eqref{eq:ODES-2ndFeq-functions-analytic-sol} is then also satisfied.

The main points of this cosmology are
\begin{itemize}   
  \item[(i)]
an expanding Friedmann-type universe with cosmic scale factor
$r \sim \tau^{1/5}$.
  \item[(ii)]
a decreasing perfect-fluid energy density and
pressure $r_{\chi}(\tau)=3\,p_{\chi}(\tau) \sim \tau^{-4/5}$.
  \item[(iii)]
a cosmological constant cancelled
by $\sqrt{-g}=\widehat{r}_\text{\,sol}$ from \eqref{eq:rhat-analytic-sol},
provided condition \eqref{eq:lambda-condition-analytic-sol} holds,
so that $r_\text{vac}(\tau)=0$.
  \item[(iv)]
the curvature scalars $\mathcal{R}(\tau) \sim 0$ and
$\mathcal{K}(\tau)\sim \tau^{-8/5}$,
approaching Minkowski spacetime for $\tau \to \infty$.  
\end{itemize} 
Observe that, for a given value $\chempot$, 
we have a whole family of asymptotic solutions  
depending on the cosmological constant $\lambda$
via \eqref{eq:rhat-analytic-sol}, as long as
condition \eqref{eq:lambda-condition-analytic-sol} holds.
The numerics must tell us whether or not
there is an attractor behavior towards Minkowski spacetime.

\subsection{Initial boundary conditions}
\label{subsec:Initial-boundary-conditions}

Before we turn to the numerical evaluation of
the ODEs \eqref{eq:dimensionless-ODEs}, we need to address
the delicate issue of boundary conditions.
We are concerned with three functions:
$a(\tau)$, $r(\tau)$, and $r_{\chi}(\tau)$.
The corresponding function space is vast and we will seek
guidance from the analytic solution obtained in
Sec.~\ref{subsec:Analytic-solution-wchi-onethird}.

So, we will start out 
at an arbitrary coordinate time $\tau_\text{bcs} > 0$ and obtain 
boundary conditions on the three functions by
considering small perturbations of the analytic solution
as given by \eqref{eq:functions-analytic-sol}, \eqref{eq:rhat-analytic-sol}, 
and \eqref{eq:constants-analytic-sol}, for $\alpha=1$.
There are two kinds of perturbations, those that do not change   
$r_\text{vac}(\tau_\text{bcs})=0$ to leading order
and those that do
(let us call the first kind ``mild'' and the second kind ``dangerous'').
We have two types of perturbations in the first category
[keeping $r_\text{vac}(\tau_\text{bcs}) \approx 0$] and
one type of perturbations in the second category
[making $r_\text{vac}(\tau_\text{bcs})$ nonzero
to leading order in the perturbation].

The first type of ``mild'' perturbations
can be written as follows:
\bsubeqs\label{eq:good-perturbations}
\beqa
\label{eq:good-perturbations-type1}
\left\{
\frac{\delta a}{a},\,
\frac{\delta r}{r},\,
\frac{\delta r_{\chi}}{r_{\chi}}
\right\}^{(\text{type-}1)}
&=&
\left\{  0,\,0,\,\delta_{1}\right\}\,,
\eeqa
with a negative or positive infinitesimal $\delta_{1}$.
For this type-1 perturbation,  the vacuum energy density
\eqref{eq:dimensionless-ODE-rvac} stays strictly zero.

The second type of ``mild'' perturbations has
\beqa
\label{eq:good-perturbations-type2}
\left\{
\frac{\delta a}{a},\,
\frac{\delta r}{r},\,
\frac{\delta r_{\chi}}{r_{\chi}}
\right\}^{(\text{type-}2)}
&=&
\left\{  
-\, 6\:\delta_{2},\,
\delta_{2},\,
0
\right\}\,,
\eeqa
with a negative or positive infinitesimal $\delta_{2}$.
The actual ratio of $\delta r/r$ and $\delta a/a$
in \eqref{eq:good-perturbations-type2} keeps the
combination $\sqrt{a}\,|r|^{3}$ unchanged
to first order in $\delta_{2}$ and precisely
this combination enters the vacuum energy density
\eqref{eq:dimensionless-ODE-rvac}.

The  third type of  perturbations is in the  ``dangerous'' category,
having $r_\text{vac} \ne 0$ 
at the starting value $\tau=\tau_\text{bcs}\,$,  
\beqa
\label{eq:bad-perturbations-type3}
\left\{
\frac{\delta a}{a},\,
\frac{\delta r}{r},\,
\frac{\delta r_{\chi}}{r_{\chi}}
\right\}^{(\text{type-}3)}
&=&
\left\{ 
\delta_{3}/6,\,
\delta_{3},\, 
0
\right\}\,,
\eeqa
\esubeqs
with a negative or positive infinitesimal $\delta_{3}$.
These perturbations give, indeed,
$r_\text{vac} \propto \delta_{3}$, which can be negative or positive.

Note, finally, that the above three perturbations
are mutually orthogonal.

\begin{figure}[t]
\vspace*{0mm}
\begin{center}
\hspace*{0mm}   
\includegraphics[width=1\textwidth]{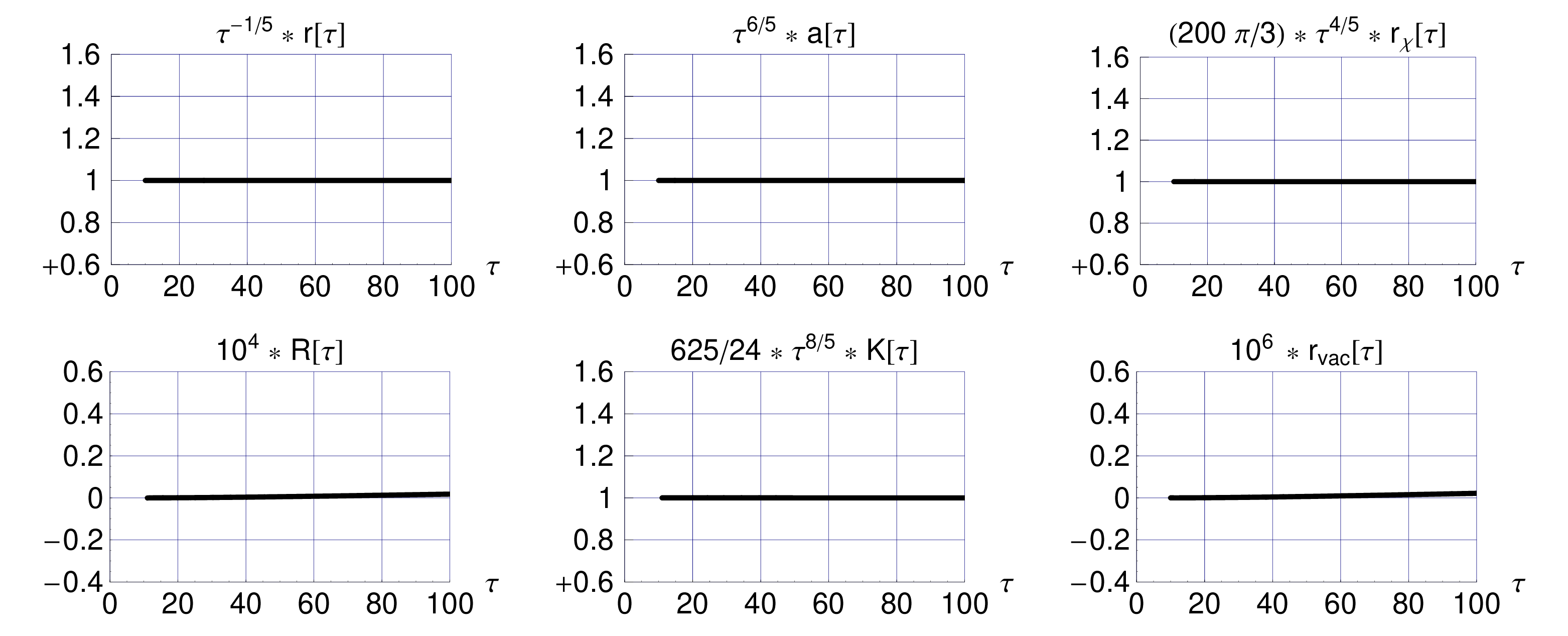}
\end{center}
\caption{Numerical solution of the ODEs \eqref{eq:dimensionless-ODEs}
with parameters $w_{M}=1/3$, $\zeta=1$, $\chempot=3$, and $\lambda=1$.
The initial boundary conditions are
taken from the analytic solution \eqref{eq:functions-analytic-sol},
\eqref{eq:rhat-analytic-sol}, and \eqref{eq:constants-analytic-sol},
for $\alpha=1$.
Specifically, the boundary conditions at $\tau=\tau_\text{bcs}=10$
are: $\{ a,r,\dot{r},r_{\chi}\}$ $=$
$\{  0.0630957344,1.58489319,0.0316978638,0.000756730758  \}$, 
where the $\dot{r}$ value has been
obtained from the first Friedman equation \eqref{eq:dimensionless-ODE-1stF}.
The top row shows the three basic variables: the metric functions
$r(\tau)$ and $a(\tau)$ 
and the dimensionless matter energy density $r_{\chi}$.  
The bottom row shows three derived quantities:
the dimensionless Ricci curvature scalar $\mathcal{R}$,
the dimensionless Kretschmann curvature scalar $\mathcal{K}$,
and the dimensionless gravitating vacuum energy density
$r_\text{vac}$ from \eqref{eq:dimensionless-ODE-rvac}.  
}
\label{fig:num-sol-lambda1-unpert}
\vspace*{0mm}
\end{figure}

\begin{figure}[t]
\vspace*{0mm}
\begin{center}
\hspace*{0mm}   
\includegraphics[width=1\textwidth]{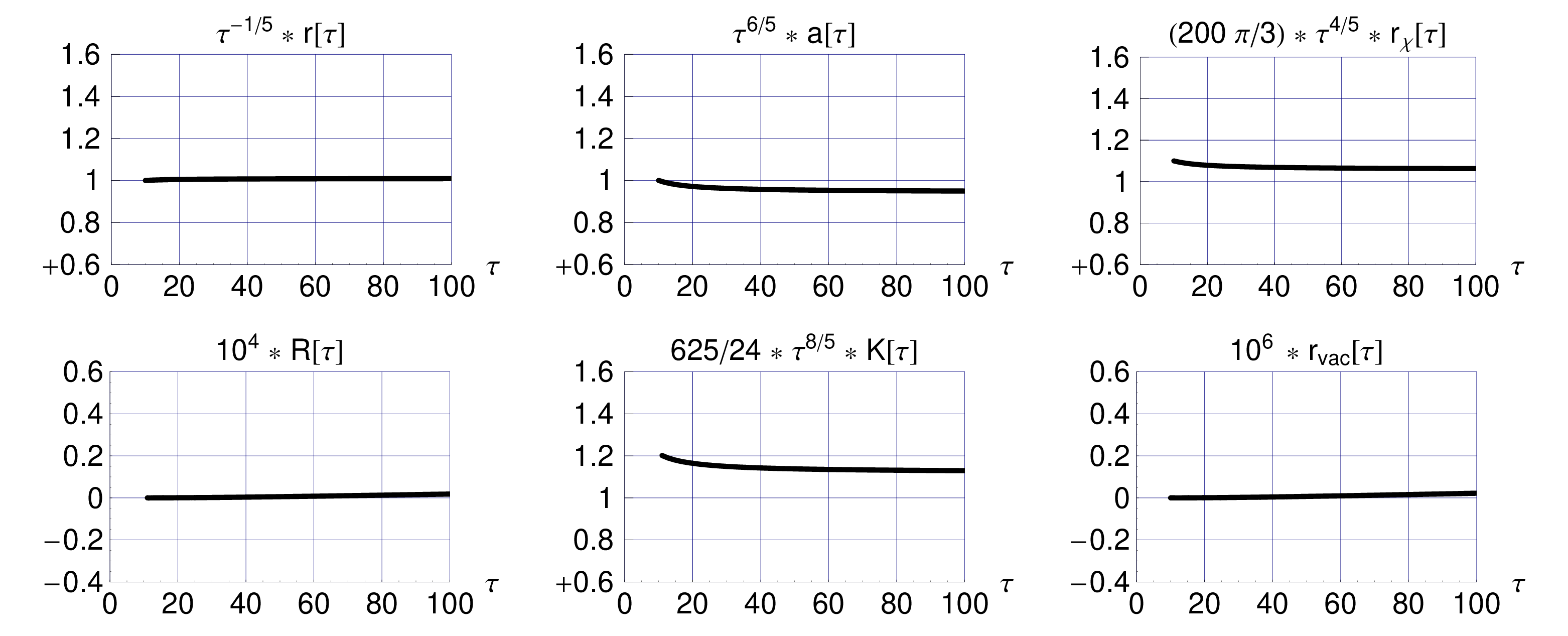}
\end{center}
\caption{The model parameters and
boundary conditions at $\tau=\tau_\text{bcs}=10$ are
as in Fig.~\ref{fig:num-sol-lambda1-unpert},
but now with a type-1 perturbation \eqref{eq:good-perturbations-type1}
for a constant $\delta_{1}=+1/10$:  
$\{ a,r,\dot{r},r_{\chi}\}$ $=$
$\{ 0.0630957344, 1.58489319, 0.0332450001, 0.000832403833 \}$,  
where the $\dot{r}$ value has been
obtained from the first Friedman equation \eqref{eq:dimensionless-ODE-1stF}.
Similar results are obtained for $\delta_{1}=-1/10$.
}
\label{fig:num-sol-lambda1-v1pert-deltaplus0pt1}
\vspace*{15mm}
\vspace*{0mm}
\begin{center}
\hspace*{0mm}
\includegraphics[width=1\textwidth]{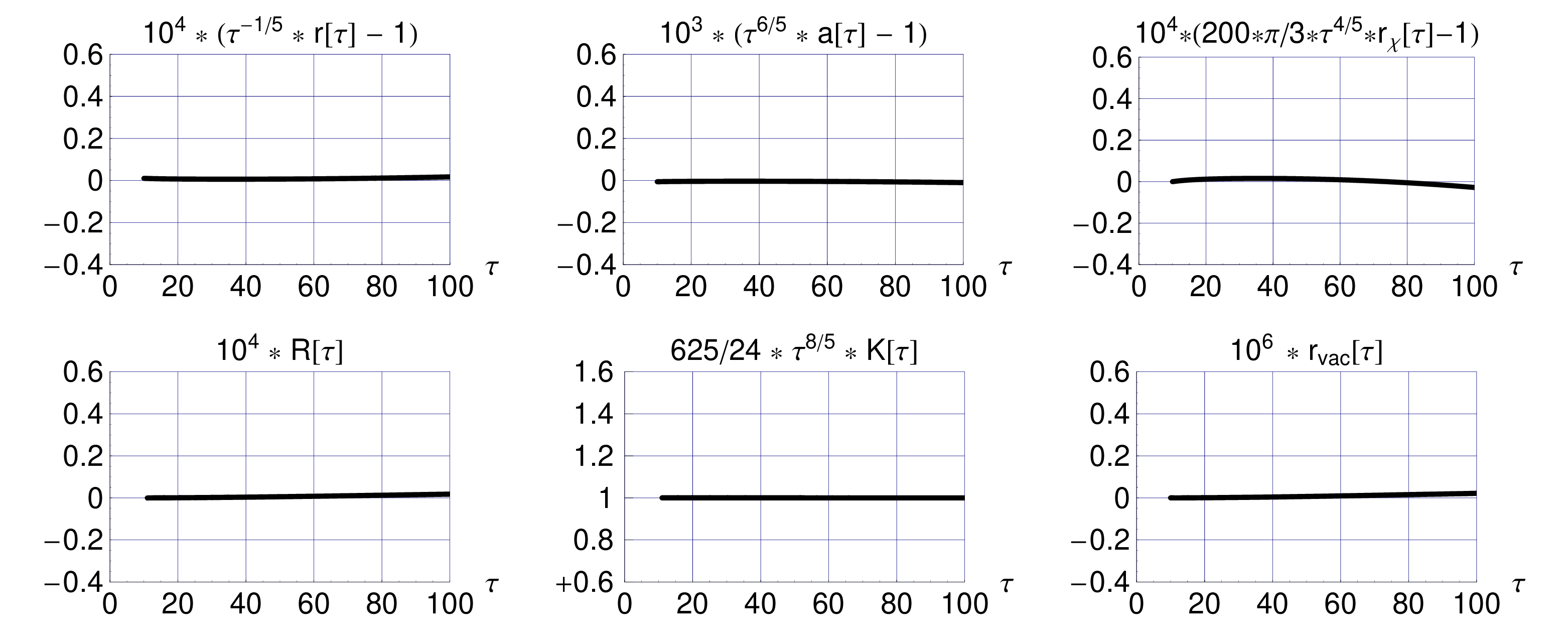}
\end{center}
\caption{The model parameters and
boundary conditions at $\tau=\tau_\text{bcs}=10$ are
as in Fig.~\ref{fig:num-sol-lambda1-unpert},
but now with a type-2 perturbation \eqref{eq:good-perturbations-type2}
for a constant $\delta_{2}=+10^{-6}$: 
$\{ a,r,\dot{r},r_{\chi}\}$ $=$
$\{ 0.0630953559, 1.58489478, 0.0316978000, 0.000756730758 \}$,   
where the $\dot{r}$ value has been
obtained from the first Friedman equation \eqref{eq:dimensionless-ODE-1stF}.
Similar results are obtained for $\delta_{2}=-10^{-6}$.
}
\label{fig:num-sol-lambda1-v2pert-deltaplus1Eminus5}
\vspace*{0mm}
\vspace*{155mm}
\end{figure}

\subsection{Numerical results}
\label{subsec:Numerical-results-original-ODES}

The ODEs \eqref{eq:dimensionless-ODEs} can
be solved numerically. Specifically,
we use the first-order ODE \eqref{eq:dimensionless-ODE-rchidoteq}
and the second-order ODE \eqref{eq:dimensionless-ODE-2ndF},
together with the $\tau$ derivative of the first-order
ODE \eqref{eq:dimensionless-ODE-1stF},
while fixing the boundary conditions
to satisfy \eqref{eq:dimensionless-ODE-1stF} as a constraint.

Numerical results for $\chempot=3$ and $\lambda=1$ are shown in
Fig.~\ref{fig:num-sol-lambda1-unpert}
for boundary conditions from the analytic solution
of Sec.~\ref{subsec:Analytic-solution-wchi-onethird}.
The numerical solution of Fig.~\ref{fig:num-sol-lambda1-unpert}
essentially reproduces the analytic solution, which
allows us to test the numerical accuracy.
In fact, we see a small error building up  
in the dimensionless Ricci curvature scalar $\mathcal{R}(\tau)$,  
but the gravitating vacuum energy density
$r_\text{vac}(\tau)$ stays close to zero within an
accuracy of $10^{-7}$.

Further numerical results are shown
in Fig.~\ref{fig:num-sol-lambda1-v1pert-deltaplus0pt1}
for boundary conditions
from a type-1 perturbation \eqref{eq:good-perturbations-type1}.
The other ``mild'' perturbation \eqref{eq:good-perturbations-type2}
gives the numerical results shown in
Fig.~\ref{fig:num-sol-lambda1-v2pert-deltaplus1Eminus5}.
The numerical solution of
Fig.~\ref{fig:num-sol-lambda1-v1pert-deltaplus0pt1}
asymptotically approaches the analytic solution
from Sec.~\ref{subsec:Analytic-solution-wchi-onethird}   
for an $\alpha$ value approximately equal to $0.95$.
The numerical solution of
Fig.~\ref{fig:num-sol-lambda1-v2pert-deltaplus1Eminus5}
also gets close to the analytic solution but not
perfectly so, as $r_\text{vac}(\tau)$ is not exactly zero
(the linear term in $\delta_2$ vanishes at $\tau=\tau_\text{bcs}$
but not the quadratic term).

For these five different boundary conditions
(unperturbed, type-1 perturbations with $\delta_{1}=\pm 1/10$,
and type-2 perturbations with $\delta_{2}=\pm 10^{-6}$),
the vacuum energy density is found to be cancelled
to high precision (less than $10^{-7}$
for the numerical solutions shown, where $\lambda=1$
sets the scale).
Obviously, these ``mild'' type-1 and type-2 perturbations
have $r_\text{vac} \approx 0$ at the starting value $\tau=\tau_\text{bcs}$,
which is then not changed by the later dynamics
[numerically, a nontrivial result; analytically,
we have $\dot{r}_\text{vac}=0$, 
as discussed in
Sec.~\ref{subsec:Cosmology-Basic-model-ODEs}].  

As mentioned in Sec.~\ref{subsec:Initial-boundary-conditions},
``dangerous'' perturbations
have $r_\text{vac} \ne 0$ at the starting value $\tau=\tau_\text{bcs}$.
We have obtained numerical results for
a type-3 perturbation \eqref{eq:bad-perturbations-type3}
with $\delta_{3}=+10^{-6}$, giving a constant vacuum energy
density $r_\text{vac}= 6.17444 \times 10^{-6}$
(a related figure will be given in
Sec.~\ref{subsec:Numerical-results-modified-ODES}).
Apparently, we need to modify the dynamics,
in order to cure the ``dangerous'' perturbations.
For this reason, modified ODEs with vacuum-matter energy exchange 
(earlier work~\cite{Klinkhamer2017} already suggested
the need for this type of energy exchange)
will be introduced in the next section.

\newpage  
\section{Cosmology: Quantum-dissipative effects}
\label{sec:Cosmology-Quantum-dissipative-effects} 

\subsection{Preliminary remarks}

A general discussion of relaxation effects in $q$-theory  
has been presented in Ref.~\cite{KlinkhamerSavelainenVolovik2016}.
A specific calculation,
for a standard spatially-flat Robertson--Walker metric
[i.e., $\widetilde{A}(t)=1$ in \eqref{eq:extRW-ds2}],
relies on particle production
by spacetime curvature~\cite{ZeldovichStarobinsky1977}.
The resulting Zeldovich--Starobinsky-type
source term reads~\cite{KlinkhamerVolovik-MPLA-2016}
\beq
\label{eq:Gamma-particle-production}
\Gamma_\text{particle-production}= 
\widehat{\gamma}\,   
\left| \widetilde{R}^{-1} \; \frac{d \widetilde{R}}{d t} \right|\,R^{2}\,,
\eeq
with the cosmic scaling function $\widetilde{R}(t)$ 
of the metric \eqref{eq:extRW-ds2}
and the Ricci curvature scalar 
$R(t)=R\big[\widetilde{A}(t),\,\widetilde{R}(t)\,\big]$.

We then have for the cosmic evolution of
the matter and vacuum energy densities:
\bsubeqs\label{eq:VM-energy-exchange}
\beqa
\label{eq:VM-energy-exchange-a}
\frac{d \rho_M}{d t} + \cdots &=& + \Gamma_\text{particle-production}\,,
\\[1mm]
\label{eq:VM-energy-exchange-b}
\frac{d \rho_V}{d t} + \cdots &=& - \Gamma_\text{particle-production}\,,
\eeqa
\esubeqs
because of energy conservation \eqref{eq:combined-energy-momentumconservation}.
Observe that Eqs. \eqref{eq:VM-energy-exchange-a} and
\eqref{eq:VM-energy-exchange-b} are 
time-reversal noninvariant for
the source term as given by  \eqref{eq:Gamma-particle-production}.
This time-reversal noninvariance is, of course, to be
expected for a dissipative effect, in fact a quantum-dissipative effect
as particle creation or annihilation is a genuine quantum
phenomenon.

\subsection{Modified ODEs with vacuum-matter energy exchange}
\label{subsec:Modified-ODEs}

We now consider a relativistic matter component with 
equation-of-state parameter $w_{M} \equiv P_{\chi}/\rho_{\chi} =1/3$
and add a positive source term $\Gamma$ on the right-hand side of
\eqref{eq:dimensionless-ODE-rchidoteq}.
We then need to
determine how this addition feeds into the other ODEs.
Specifically, we take three steps towards modified ODEs with a
phenomenological implementation of quantum-dissipative effects.
Henceforth, we use the dimensionless variables  
from \eqref{eq:dimensionless-variables}.

In step 1, we add a source term $\Gamma$ to the right-hand side
of \eqref{eq:dimensionless-ODE-rchidoteq} for $w_{M}=1/3$ to get 
\bsubeqs\label{eq:dimensionless-modified-ODEs}
\beqa
\label{eq:dimensionless-modified-ODEs-eq1}
\hspace*{0mm}
&&
\dot{r}_{\chi} + 4\,
\left(\frac{\dot{r}}{r}\right)\,r_{\chi} = \Gamma \,.
 \eeqa
Next, we see how the new term in
\eqref{eq:dimensionless-modified-ODEs-eq1}  feeds
into the two ODEs \eqref{eq:dimensionless-ODE-1stF}
and \eqref{eq:dimensionless-ODE-2ndF}.

In step 2, we eliminate $r_{\chi}$
by taking the sum of one third of
\eqref{eq:dimensionless-ODE-1stF}
and \eqref{eq:dimensionless-ODE-2ndF} for $w_{M}=1/3$,  
\beqa
\label{eq:dimensionless-modified-ODEs-eq2} 
&&
\frac{1}{8\,\pi \,a}\,
\left[
\frac{2\,\ddot{r}}{r} +2\,
\left( \frac{\dot{r}}{r} \right)^{2}
-\left( \frac{\dot{r}}{r} \right)\,\left( \frac{\dot{a}}{a} \right)
\right] =
\frac{4}{3}\,
r_\text{vac}\,,
\eeqa
where the $r_\text{vac}$ expression will be recalled shortly.

In step 3,   
we take the derivative of \eqref{eq:dimensionless-ODE-1stF},
use \eqref{eq:dimensionless-modified-ODEs-eq1} to eliminate $\dot{r}_{\chi}$,
use \eqref{eq:dimensionless-ODE-1stF} to eliminate $r_{\chi}$,
use the $\ddot{r}$ expression from \eqref{eq:dimensionless-modified-ODEs-eq2},
and get
\beqa
\label{eq:dimensionless-modified-ODEs-eq3} 
\hspace*{0mm}
&&
\dot{r}_\text{vac}=-\Gamma\,,
\\[2.0mm]
\label{eq:dimensionless-modified-ODEs-rvac}
\hspace*{0mm}
&&
r_\text{vac}  =
\lambda   + 2\,\zeta\,\sqrt{a}\,|r|^{3}-\chempot \,,
\eeqa
\esubeqs
where the explicit $r_\text{vac}$ expression has now  been repeated.
For completeness, we give the original first-order Friedman equation,
\beqa
\label{eq:dimensionless-modified-ODEs-1stFeq}
\hspace*{0mm}
&&
3\,\left( \frac{\dot{r}}{r} \right)^{2} =
8\,\pi \,a\, \Big(r_{\chi}+r_\text{vac} \Big)\,,
\eeqa
which, if it holds initially for the
solution of the ODEs \eqref{eq:dimensionless-modified-ODEs},
will be satisfied at subsequent times (later on, 
this will make for a useful diagnostic of the numerical accuracy).

Two remarks are in order:
\begin{itemize}
  \item[(1)]
For Minkowski spacetime with $a(\tau)=r(\tau)=1$
in the dimensionless version of \eqref{eq:extRW-ds2},
we have $\dot{r}_{\chi} = \Gamma$ from \eqref{eq:dimensionless-modified-ODEs-eq1}
and
$\dot{r}_\text{vac} = - \Gamma$ from \eqref{eq:dimensionless-modified-ODEs-eq3},
which corresponds to
a direct vacuum-matter energy exchange as long as $\Gamma$ is nonvanishing.
  \item[(2)]
It would appear that the ODEs  
\eqref{eq:dimensionless-modified-ODEs-eq2}and \eqref{eq:dimensionless-modified-ODEs-eq3}
are independent of the matter energy density $r_{\chi}$,
but that dependence enters by the use of the first-order
Friedman equation \eqref{eq:dimensionless-modified-ODEs-1stFeq}
as a constraint on the boundary conditions.
\end{itemize}

Another point is the choice of $\Gamma$
so that the numerics works.
A suitable choice is
\bsubeqs\label{eq:Gamma-tilde}
\beqa
\Gamma &=& \widetilde{\gamma}\;
|\dot{r}/r|\;\left( r_\text{vac} \right)^{2}\,,
\\[2.0mm]
\widetilde{\gamma}(\tau) &=& \gamma\;
\left[
\frac{\tau^{2}-\tau_\text{bcs}^{2}}{\tau^{2}+1}
\right]^{2}\,,
\\[2.0mm]
\gamma &\geq& 0\,,
\eeqa
\esubeqs
for initial boundary conditions at  $\tau=\tau_\text{bcs}$.
This basically has the structure of
expression \eqref{eq:Gamma-particle-production},
because the left-hand side
of \eqref{eq:dimensionless-modified-ODEs-eq2}is proportional to the Ricci scalar
[recall that we have a matter component with $w_{M}=1/3$,
so that the  right-hand side
of \eqref{eq:dimensionless-modified-ODEs-eq2}vanishes if there is no vacuum component].
We have added in \eqref{eq:Gamma-tilde}
a smooth switch-on function
$\widetilde{\gamma}(\tau)$, in order to ease the
numerical evaluation of the ODEs.

\begin{figure}[t]
\vspace*{0mm}
\begin{center}
\hspace*{0mm}
\includegraphics[width=1\textwidth]{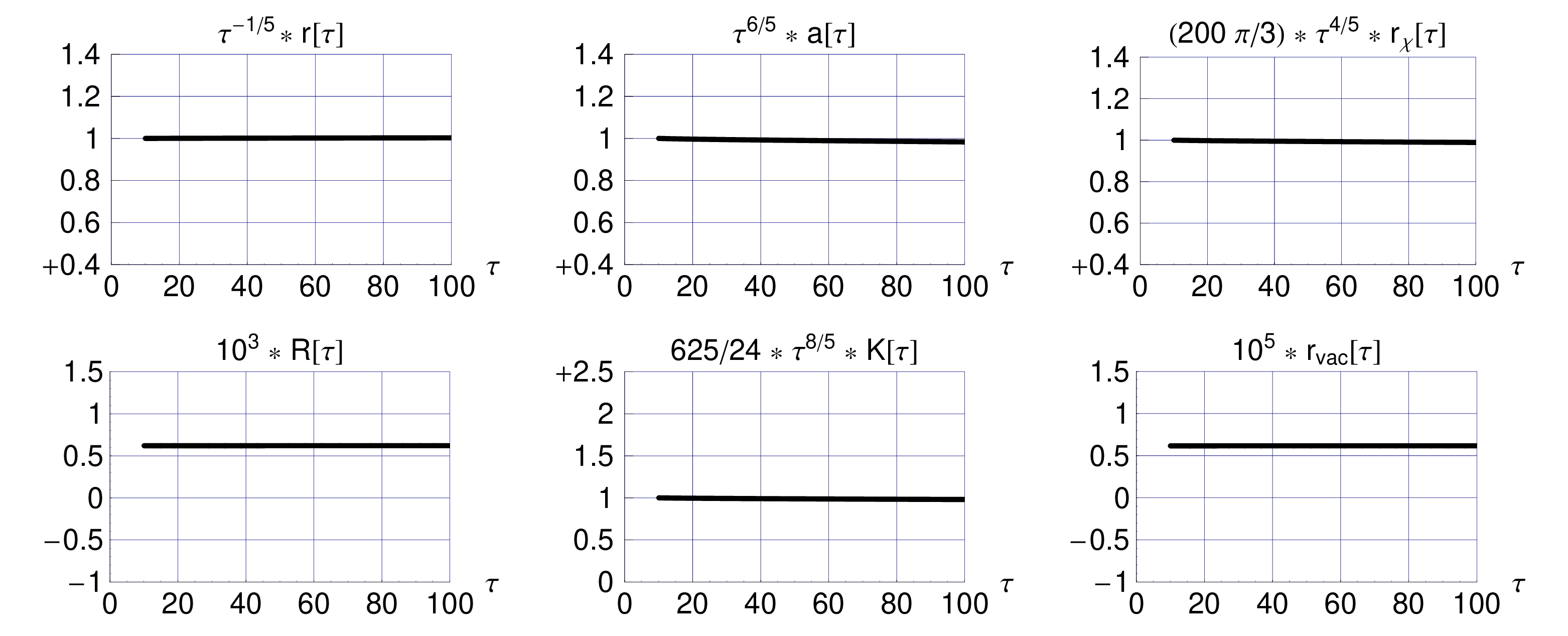}
\end{center}
\caption{Numerical solution of the modified
ODEs \eqref{eq:dimensionless-modified-ODEs} with
source term \eqref{eq:Gamma-tilde}
and parameters $w_{M}=1/3$,
$\zeta=1$, $\chempot=3$, $\lambda=1$, and $\gamma=0$
(quantum-dissipative effects turned off).
The initial boundary conditions are
taken from the analytic solution \eqref{eq:functions-analytic-sol},
\eqref{eq:rhat-analytic-sol}, and \eqref{eq:constants-analytic-sol},
for $\alpha=1$,
but now with a type-3 perturbation \eqref{eq:bad-perturbations-type3}
for $\delta_{3}=+10^{-6}$.
Specifically, the boundary conditions at $\tau=\tau_\text{bcs}=10$ are:
$\{ a,r,\dot{r},r_{\chi}\}$ $=$
$\{   0.0630957450, 1.58489478, 0.03182679077, 0.000756730758 \}$, 
where the $\dot{r}$ value has been obtained from
the first Friedman equation \eqref{eq:dimensionless-modified-ODEs-1stFeq}.
The top row shows the three basic variables: the metric functions
$r(\tau)$ and $a(\tau)$ 
and the dimensionless matter energy density $r_{\chi}$.  
The bottom row shows three derived quantities:
the dimensionless Ricci curvature scalar $\mathcal{R}$,
the dimensionless Kretschmann curvature scalar $\mathcal{K}$,
and the dimensionless gravitating vacuum energy density
$r_\text{vac}$ from \eqref{eq:dimensionless-modified-ODEs-rvac}.
The vacuum energy density from the initial conditions
is $r_\text{vac}(\tau_\text{bcs})=6.16667\times 10^{-6}$,
which stays essentially constant.
}
\label{fig:num-sol-lambda1-v3pert-deltaplus1Emin6-gamma0-modODEs}
\end{figure}

Observe, again, that the
ODEs \eqref{eq:dimensionless-modified-ODEs-eq1}
and \eqref{eq:dimensionless-modified-ODEs-eq3}
with source term \eqref{eq:Gamma-tilde} are time-reversal noninvariant.
The basic structure of the resulting vacuum-energy equation,
\beq
\label{eq:dimensionless-modified-ODEs-rvacdoteq-structure}
 \dot{r}_\text{vac}
 = - \widetilde{\gamma}\;
\big|\dot{r}/r\big|\;\big( r_\text{vac} \big)^{2}\,,
\eeq
is similar to the one discussed in
Refs.~\cite{KlinkhamerVolovik-MPLA-2016,Klinkhamer2022-preprint},
where, with a rapid switch-on,
an analytic solution could be obtained for
$r_\text{vac}(\tau)$ that drops to zero as $\tau\to\infty$.

The exact solution of Sec.~\ref{subsec:Analytic-solution-wchi-onethird}
carries over to
the modified ODEs \eqref{eq:dimensionless-modified-ODEs} 
with source term \eqref{eq:Gamma-tilde} .
The reason is simply that this source term $\Gamma$ vanishes if
$r_\text{vac}=0$, which is precisely the case for our
analytic solution.

\subsection{Numerical results}
\label{subsec:Numerical-results-modified-ODES}

Numerical results from the original ODEs 
\eqref{eq:dimensionless-ODEs}
for ``mild'' perturbations (keeping
$r_\text{vac} \approx 0$) have been discussed in
Sec~\ref{subsec:Numerical-results-original-ODES}.
These results are essentially unchanged if we use
the modified ODEs \eqref{eq:dimensionless-modified-ODEs},
as the source term \eqref{eq:Gamma-tilde}
vanishes if $r_\text{vac}$ does.
We then turn to the ``dangerous'' perturbations
(making for nonzero $r_\text{vac}$ to leading order
in the perturbation), for which
we have introduced the modified ODEs \eqref{eq:dimensionless-modified-ODEs}
with the source term $\Gamma$ from \eqref{eq:Gamma-tilde}.

\begin{figure}[b]
\begin{center}
\hspace*{0mm}
\includegraphics[width=1\textwidth]{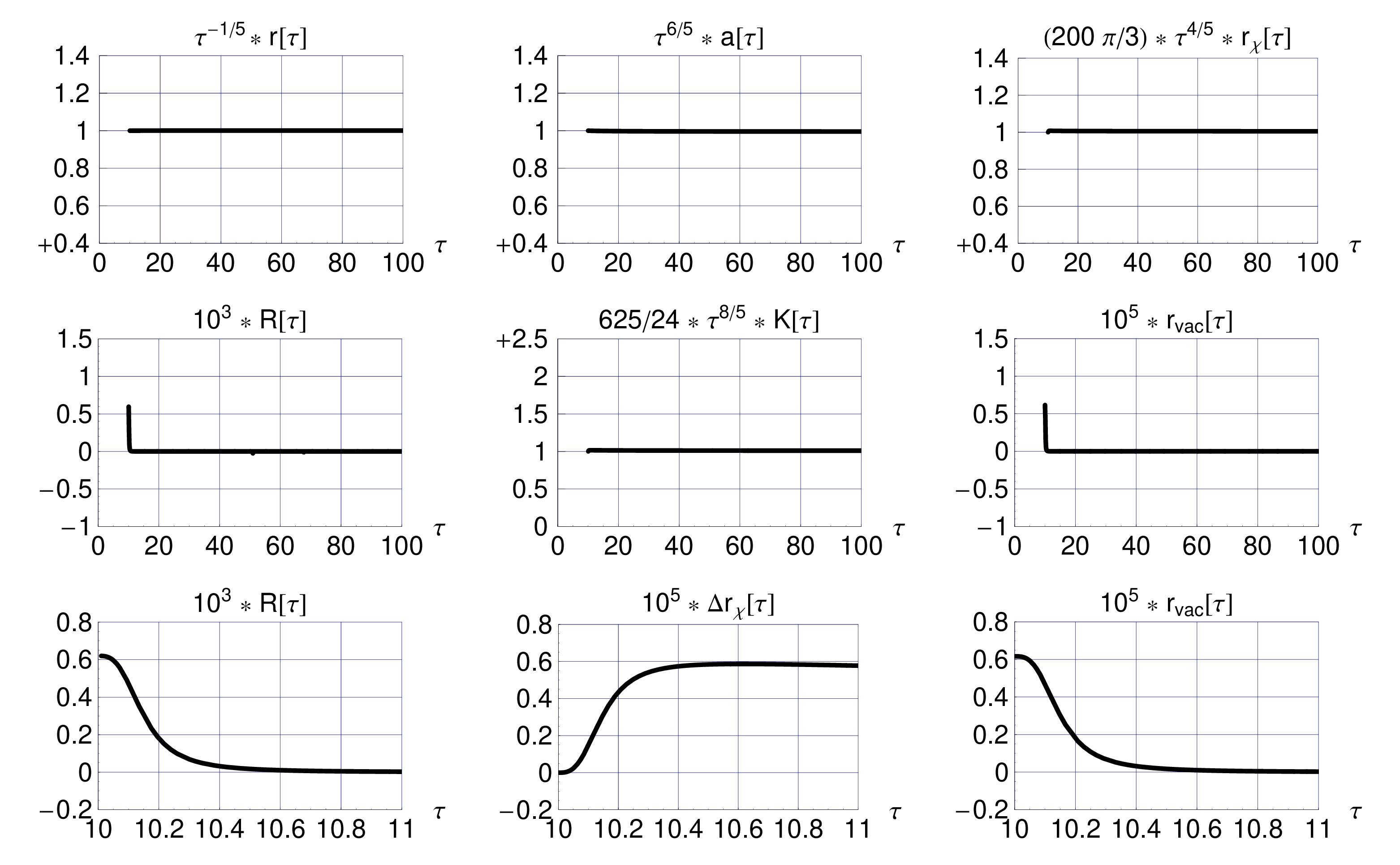}
\end{center}
\caption{The boundary conditions at $\tau=\tau_\text{bcs}=10$
and the model parameters are the same as in
Fig.~\ref{fig:num-sol-lambda1-v3pert-deltaplus1Emin6-gamma0-modODEs},
but now with $\gamma=2 \times 10^{11}$ (quantum-dissipative effects turned on).
The vacuum energy density 
is initially $r_\text{vac}(10)=6.16667\times 10^{-6}$
and drops to $r_\text{vac}(100)\sim 1 \times 10^{-10}$.  
The bottom-middle panel shows the extra contribution 
to the matter energy density compared to that of
the $\gamma=0$ numerical solution of
Fig.~\ref{fig:num-sol-lambda1-v3pert-deltaplus1Emin6-gamma0-modODEs}, 
specifically $\Delta r_{\chi}(\tau) \equiv r_{\chi}(\tau) -
r_{\chi}^{(\gamma=0,\,\text{num-sol})}(\tau)$. 
}
\label{fig:num-sol-lambda1-v3pert-deltaplus1Emin6-gamma2E11-modODEs}
\vspace*{0mm}
\end{figure}
\begin{figure}[t]
\begin{center}
\hspace*{0mm}
\includegraphics[width=1\textwidth]{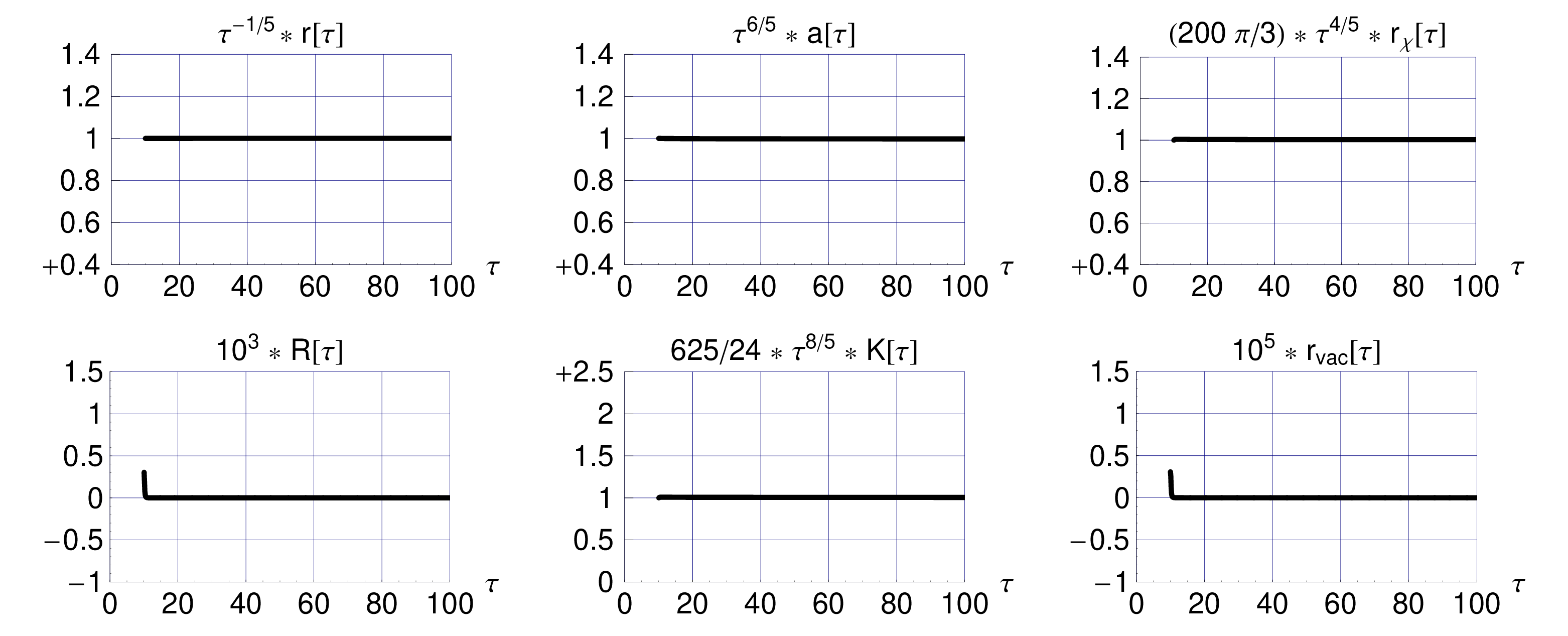}
\end{center}
\caption{The model parameters and the boundary conditions at $\tau=\tau_\text{bcs}=10$
are the same as in
Fig.~\ref{fig:num-sol-lambda1-v3pert-deltaplus1Emin6-gamma2E11-modODEs},
but now with a type-3 perturbation \eqref{eq:bad-perturbations-type3}
for $\delta_{3}=+5\times10^{-7}$:
$\{ a,r,\dot{r},r_{\chi}\}$ $=$
$\{  0.0630957397, 1.584893985, 0.03176239262, 0.000756730758   \}$, 
where the $\dot{r}$ value has been obtained from
the first Friedman equation \eqref{eq:dimensionless-modified-ODEs-1stFeq}.
The vacuum energy density 
is initially $r_\text{vac}(10)=3.08333 \times 10^{-6}$
and drops to $r_\text{vac}(100)\sim 1 \times 10^{-10}$.  
}
\label{fig:num-sol-lambda1-v3plus-delta5Emin7-gamma2E12-modODEs}
\vspace*{0mm}
\vspace*{15mm}
\vspace*{-5mm}
\begin{center}
\hspace*{0mm}
\includegraphics[width=1\textwidth]{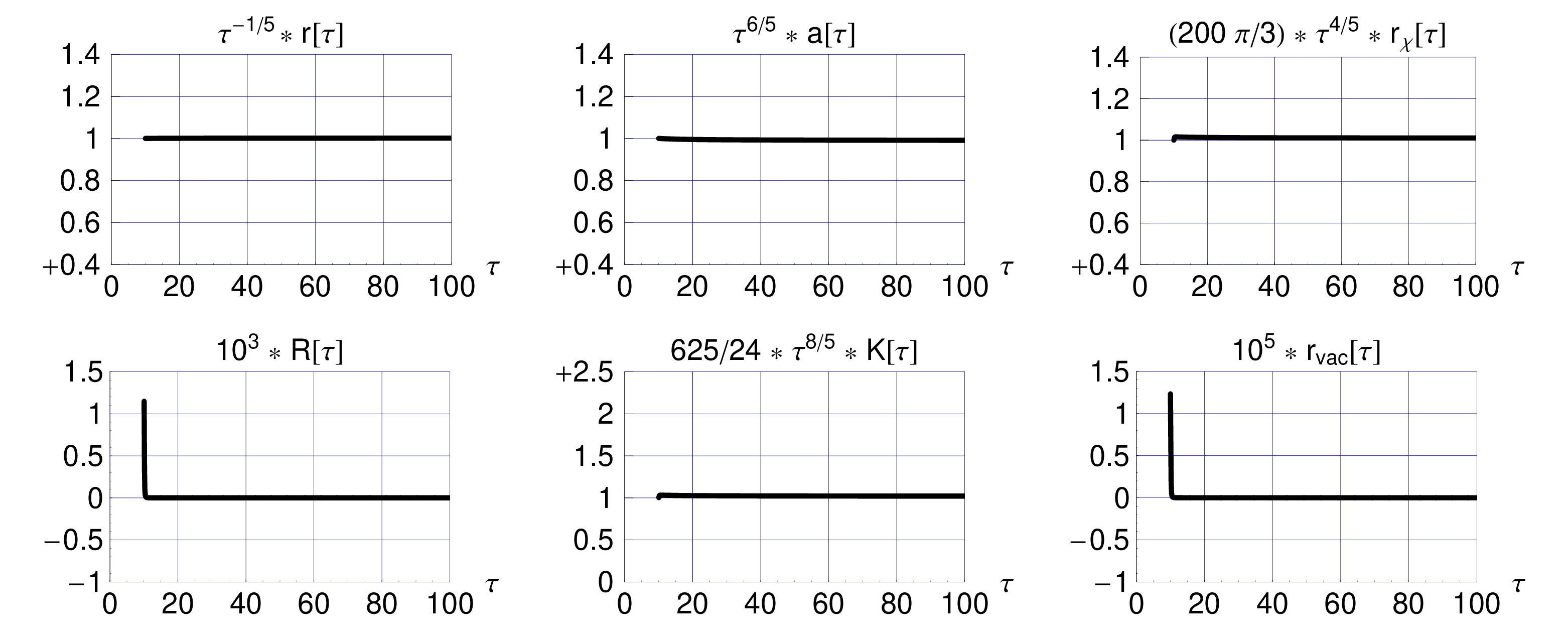}
\end{center}
\caption{The model parameters and the boundary conditions at $\tau=\tau_\text{bcs}=10$
are the same as in
Fig.~\ref{fig:num-sol-lambda1-v3pert-deltaplus1Emin6-gamma2E11-modODEs},
but now with a type-3 perturbation \eqref{eq:bad-perturbations-type3}
for $\delta_{3}=+2\times10^{-6}$:
$\{  0.0630957555, 1.584896362, 0.03195519835, 0.000756730758  \}$, 
where the $\dot{r}$ value has been obtained from
the first Friedman equation \eqref{eq:dimensionless-modified-ODEs-1stFeq}.
The vacuum energy density
is initially $r_\text{vac}(10)=1.23334\times 10^{-5}$
and drops to $r_\text{vac}(100)\sim 2 \times 10^{-10}$.
}
\label{fig:num-sol-lambda1-v3pert-deltaplus2Emin6-gamma2E11-modODEs}
\vspace*{0mm}
\end{figure}
\begin{figure}[t] 
\begin{center}
\hspace*{0mm}
\includegraphics[width=1\textwidth]{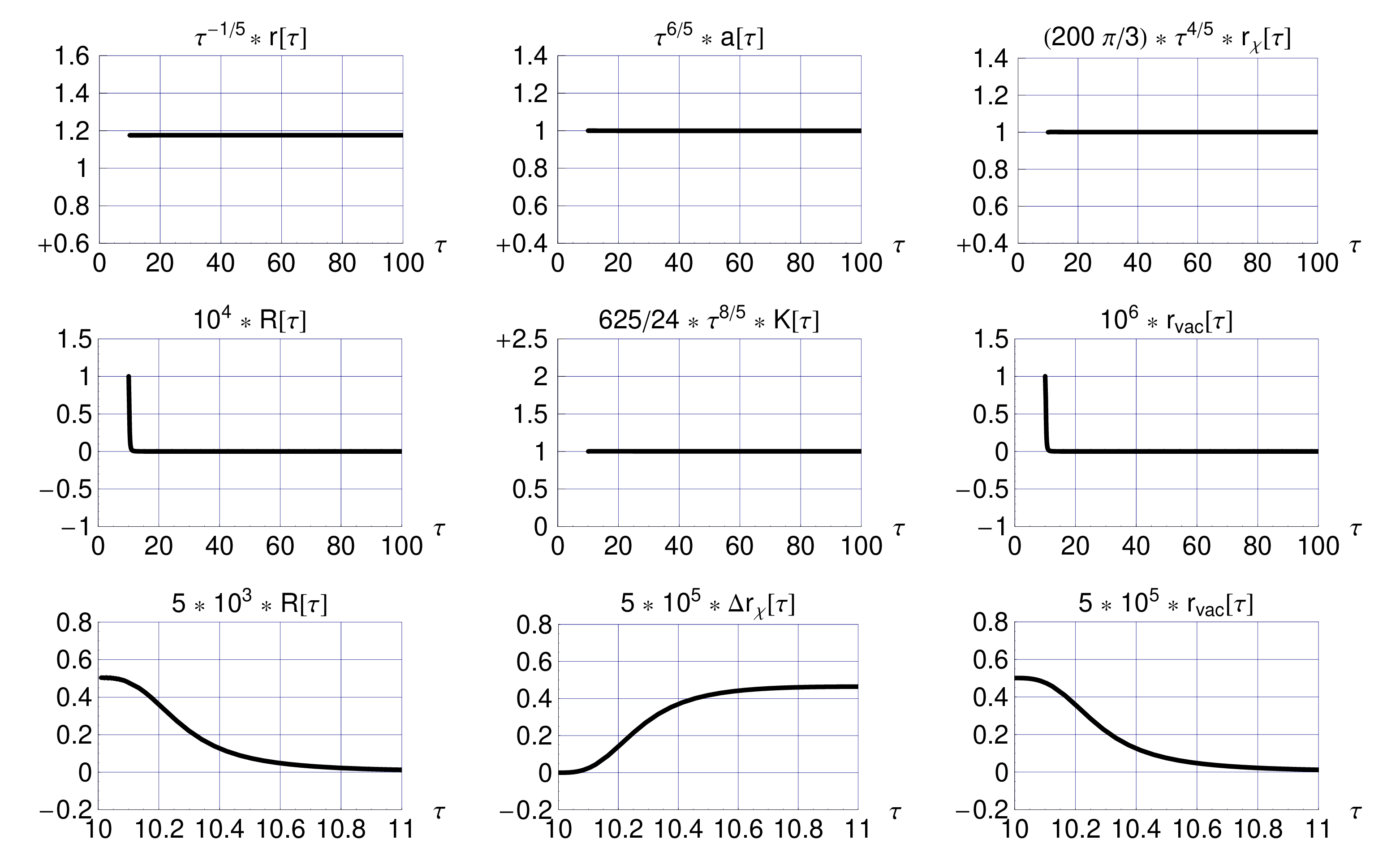}
\end{center}\vspace*{-4mm}
\caption{Numerical solution of the modified
ODEs \eqref{eq:dimensionless-modified-ODEs} with
source term \eqref{eq:Gamma-tilde} and parameters $w_{M}=1/3$,
$\zeta=1$, $\chempot=3$, $\gamma=2 \times 10^{11}$,
and $\lambda=-1/4$ (different from the value $\lambda=1$ in
Fig.~\ref{fig:num-sol-lambda1-v3pert-deltaplus1Emin6-gamma2E11-modODEs}).
The initial boundary conditions are
taken from the analytic solution \eqref{eq:functions-analytic-sol},
\eqref{eq:rhat-analytic-sol}, and \eqref{eq:constants-analytic-sol},
for $\alpha=1$,
but now with a type-3 perturbation \eqref{eq:bad-perturbations-type3}
for $\delta_{3}=+10^{-7}$.
Specifically, the boundary conditions at $\tau=\tau_\text{bcs}=10$ are:
$\{ a,r,\dot{r},r_{\chi}\}$ $=$
$\{   0.0630957355, 1.86330736,  0.0372908137,  0.000756730758 \}$,   
where the $\dot{r}$ value has been obtained from
the first Friedman equation \eqref{eq:dimensionless-modified-ODEs-1stFeq}.
The basic variables shown have been explained in the caption of
Fig.~\ref{fig:num-sol-lambda1-v3pert-deltaplus1Emin6-gamma0-modODEs}
and $\Delta r_{\chi}$ from the bottom-middle panel 
is defined in the caption of
Fig.~\ref{fig:num-sol-lambda1-v3pert-deltaplus1Emin6-gamma2E11-modODEs}. 
The vacuum energy density is initially
$r_\text{vac}(10)=1.00208\times 10^{-6}$ 
and drops to $r_\text{vac}(100)\sim 2 \times 10^{-10}$.
}
\label{fig:num-sol-lambdaminus1quarter-v3pert-deltaplus1Emin7-gamma2E11-modODEs}
\end{figure}

Numerical results, for $\chempot=3$ and $\lambda=1$, are presented
in Figs.~\ref{fig:num-sol-lambda1-v3pert-deltaplus1Emin6-gamma0-modODEs}
and \ref{fig:num-sol-lambda1-v3pert-deltaplus1Emin6-gamma2E11-modODEs}
for boundary conditions from a type-3
perturbation \eqref{eq:bad-perturbations-type3}
with $\delta_{3}=+10^{-6}$ at $\tau=\tau_\text{bcs}=10$
and two values of the 
vacuum-matter energy-exchange 
coupling constant, $\gamma=0$ and $\gamma=2 \times 10^{11}$.
Focussing on the $r_\text{vac}$ panels
in Fig.~\ref{fig:num-sol-lambda1-v3pert-deltaplus1Emin6-gamma2E11-modODEs},
we see that the modified ODEs \eqref{eq:dimensionless-modified-ODEs}
can cancel an initial positive vacuum energy density
and get an asymptotic behavior close to that of the analytic Friedmann-type
solution of Sec.~\ref{subsec:Analytic-solution-wchi-onethird}
for $\alpha \approx 0.995$.
A further remark on these $r_\text{vac}$ panels
is that the drop of $r_\text{vac}(\tau)$ by about an order of magnitude
occurs smoothly but rapidly, over the interval $\tau \in [10,\,10.3]$.
A similar drop occurs for the Ricci curvature scalar
(see the bottom-left panel of
Fig.~\ref{fig:num-sol-lambda1-v3pert-deltaplus1Emin6-gamma2E11-modODEs}),
which is consistent with the
result $\mathcal{R} =32\,\pi\,r_\text{vac} \approx 100.53\,r_\text{vac}$
from the ODE \eqref{eq:dimensionless-modified-ODEs-eq2}.
Corresponding to the drop of the vacuum energy density
(bottom-right panel of
Fig.~\ref{fig:num-sol-lambda1-v3pert-deltaplus1Emin6-gamma2E11-modODEs}),
there is an increase of the matter energy density
(bottom-middle panel), but the match between both panels is not perfect, 
which may be due to  the nonlinearity of the ODEs and the numerical accuracy.

For the same source term \eqref{eq:Gamma-tilde}
with $\gamma=2 \times 10^{11}$,
different type-3 boundary conditions also give a relaxation
to vanishing $r_\text{vac}$
(see Figs.~\ref{fig:num-sol-lambda1-v3plus-delta5Emin7-gamma2E12-modODEs}
and \ref{fig:num-sol-lambda1-v3pert-deltaplus2Emin6-gamma2E11-modODEs}).
In short, we get, for $\chempot=3$ and $\lambda=1$, a vanishing
vacuum energy density $r_\text{vac}$
from a finite domain of initial conditions,
namely $\delta_{3} \in [5\times10^{-7},\,2\times10^{-6}]$
for type-3 perturbations.
Recall, that we also have finite domains for
the type-1 and type-2 perturbations discussed
in Sec.~\ref{subsec:Numerical-results-original-ODES}.
There is, in fact, a finite 3-volume in the    
$\{ a(\tau_\text{bcs}),\,r(\tau_\text{bcs}),\,r_{\chi}(\tau_\text{bcs})\}$
space (parametrized by $\delta_{1} \in [-0.1,\,+0.1]$,
$\delta_{2} \in [-10^{-6},\,+10^{-6}]$, and
$\delta_{3} \in [5\times10^{-7},\,2\times10^{-6}]$),
whose corresponding solutions have
$r_\text{vac}=\text{O}\left(10^{-5}\right)$ initially
and $r_\text{vac}=\text{O}\left(10^{-10}\right)$ asymptotically.

Similar results have been obtained for
other values of the cosmological constant,   
provided condition \eqref{eq:lambda-condition-analytic-sol} holds.
Numerical results for $\chempot=3$ and $\lambda=-1/4$ are presented in   
Fig.~\ref{fig:num-sol-lambdaminus1quarter-v3pert-deltaplus1Emin7-gamma2E11-modODEs}.
The numerical solution of
Fig.~\ref{fig:num-sol-lambdaminus1quarter-v3pert-deltaplus1Emin7-gamma2E11-modODEs}
approaches asymptotically the analytic solution
from Sec.~\ref{subsec:Analytic-solution-wchi-onethird}
for $\widehat{r}_\text{\,sol}=(13/8)^{\,1/3}\approx 1.176$ and an $\alpha$ value
approximately equal to  $0.9992$.

\begin{figure}[b]
\vspace*{0mm}
\begin{center}
\hspace*{0mm}    
\includegraphics[width=1\textwidth]{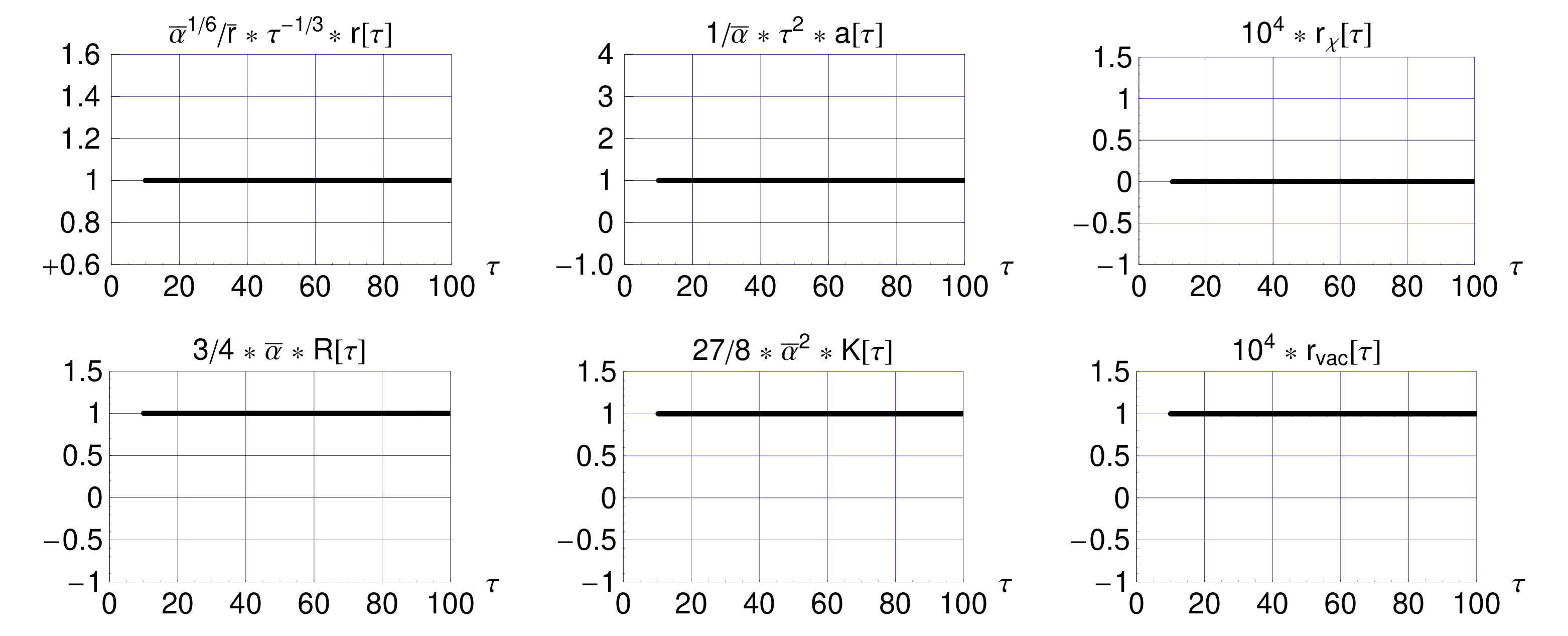}
\end{center}
\caption{Numerical solution of the modified
ODEs \eqref{eq:dimensionless-modified-ODEs} with
source term \eqref{eq:Gamma-tilde} and parameters $w_{M}=1/3$,
$\zeta=1$, $\chempot=3$, $\lambda=10^{-4}$, and $\gamma=0$
(quantum-dissipative effects turned off).
The initial boundary conditions are
taken from the analytic de-Sitter-type 
solution \eqref{eq:functions-analytic-deS-sol}
and \eqref{eq:constants-analytic-deS-spec-sol},
having $\overline{\alpha}\equiv \alpha_\text{deS-spec-sol}=132.629$ 
and $\overline{r} \equiv r_\text{deS-spec-sol}=1.14471$. 
Specifically, the boundary conditions at $\tau=\tau_\text{bcs}=10$ are:
$\{ a,r,\dot{r},r_{\chi}\}$ $=$
$\{  1.32629119, 1.09208709, 0.0364029029, 0 \}$,
where the $\dot{r}$ value has been obtained from
the first Friedman equation \eqref{eq:dimensionless-modified-ODEs-1stFeq}.
The top row shows the three basic variables: the metric functions
$r(\tau)$ and $a(\tau)$ 
and the dimensionless matter energy density $r_{\chi}$.
The bottom row shows three derived quantities: 
the dimensionless Ricci curvature scalar $\mathcal{R}$,
the dimensionless Kretschmann curvature scalar $\mathcal{K}$,
and the dimensionless gravitating vacuum energy density
$r_\text{vac}$ from \eqref{eq:dimensionless-modified-ODEs-rvac}. 
The vacuum energy density from the initial conditions
is $r_\text{vac}(\tau_\text{bcs})=1 \times  10^{-4}$,
which stays essentially constant.
}
\label{fig:num-sol-lambda0pt0001-deSbcs-gamma0-modODEs}
\vspace*{0mm}
\end{figure}

\begin{figure}[t]
\begin{center}  
\hspace*{0mm}
\includegraphics[width=1\textwidth]{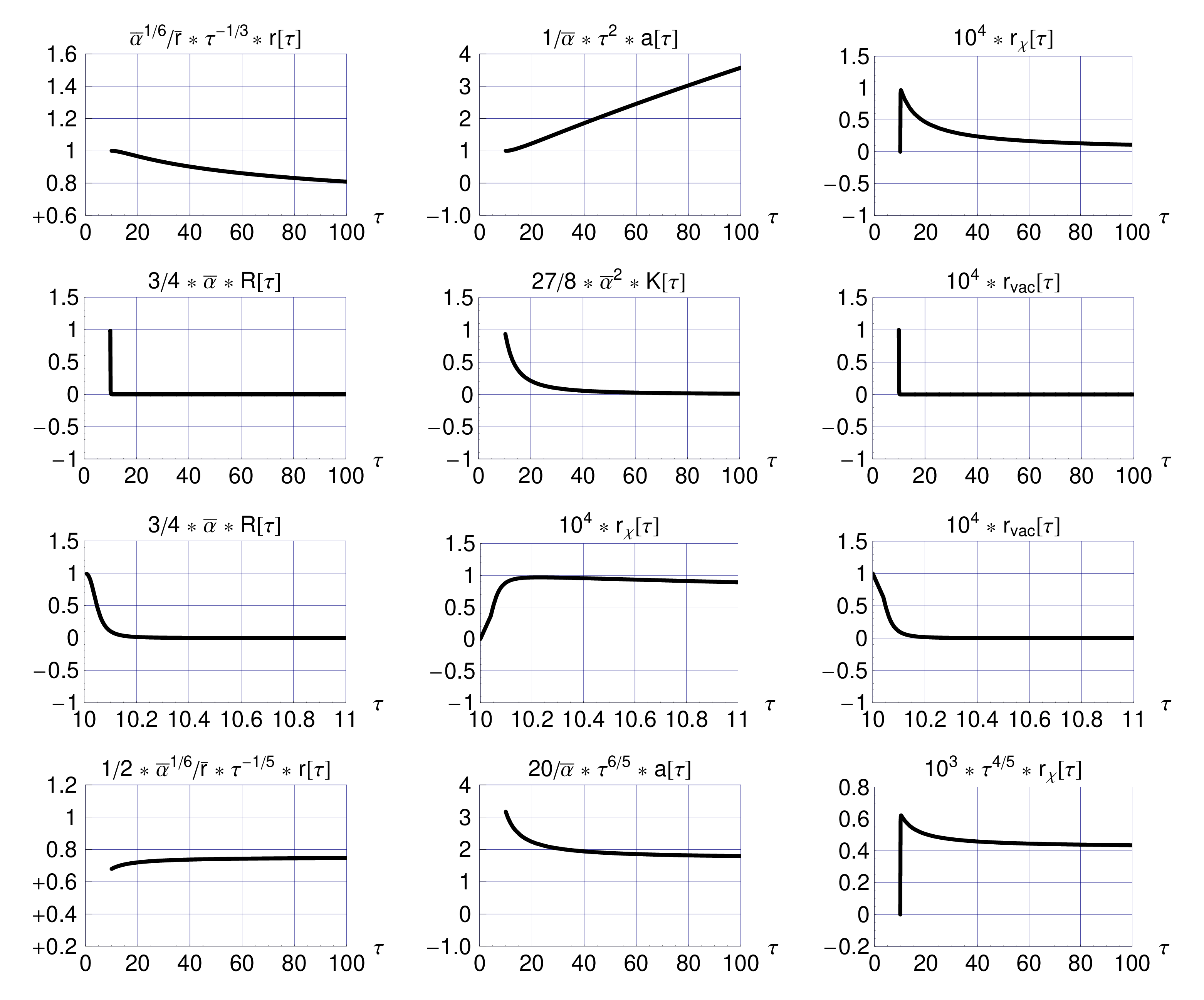}
\end{center}
\caption{The boundary conditions at $\tau=\tau_\text{bcs}=10$
and the model parameters are the same as in
Fig.~\ref{fig:num-sol-lambda0pt0001-deSbcs-gamma0-modODEs},
but now with $\gamma=2 \times 10^{11}$ (quantum-dissipative effects turned on).
The vacuum energy density 
is  initially $r_\text{vac}(10)=1 \times 10^{-4}$ 
and drops to $r_\text{vac}(100)\sim 2 \times 10^{-10}$.
The third row shows the behavior near the initial boundary conditions
and the fourth row the asymptotic behavior.
}
\label{fig:num-sol-lambda0pt0001-deSbcs-gamma2E11-modODEs}
\vspace*{0mm}
\vspace*{155mm}
\end{figure}

The previous results start ``close'' to the
analytic Friedmann-type solution in configuration space, but it is 
also possible to start ``further away'' in configuration space.
Specifically,
we can start from the analytic de-Sitter-type configuration as given
in App.~\ref{app:Analytic-solution-deS}.
Numerical results, for $\chempot=3$ and $\lambda=10^{-4}$, are presented
in Figs.~\ref{fig:num-sol-lambda0pt0001-deSbcs-gamma0-modODEs}
and \ref{fig:num-sol-lambda0pt0001-deSbcs-gamma2E11-modODEs}
with two values of the vacuum-matter-energy-exchange coupling constant $\gamma$.
The numerical solution of Fig.~\ref{fig:num-sol-lambda0pt0001-deSbcs-gamma0-modODEs}
with $\gamma=0$ essentially reproduces the analytic solution of
App.~\ref{app:Analytic-solution-deS}, whereas
the numerical solution 
of Fig.~\ref{fig:num-sol-lambda0pt0001-deSbcs-gamma2E11-modODEs}
with $\gamma=2 \times 10^{11}$
shows the reduction of the vacuum energy density $r_\text{vac}$
and the approach to the analytic Friedmann-type solution
of Sec.~\ref{subsec:Analytic-solution-wchi-onethird}
[see, in particular, the bottom-row panels
in Fig.~\ref{fig:num-sol-lambda0pt0001-deSbcs-gamma2E11-modODEs}
with $r(\tau) \propto \tau^{1/5}$,  $a(\tau)  \propto \tau^{-6/5}$, 
and $r_{\chi}(\tau)  \propto \tau^{-4/5}$\,].

To summarize, it has been shown that
the cosmological constant $\Lambda$ can, in principle,
be cancelled by $\sqrt{-g}$
and appropriate quantum-dissipative effects. For completeness, 
we give, in App.~\ref{app:Readjustment-after-PhT},  
further numerical results on how the vacuum energy density
is cancelled before and after a phase transition,
making concrete the general remarks 
in the last paragraph 
of Sec.~\ref{sec:Metric-determinant-as-dynamic-variable}.

\newpage   
\section{Final remarks}
\label{sec:Final-remarks}

Perhaps the most interesting suggestion of this paper is the interpretation
of the action \eqref{SN} for $n(x) \propto \sqrt{-g(x)}$,
as discussed in Sec.~\ref{sec:Crystal}.
In the standard formulation of general relativity,
this action is just a cosmological constant term
with ``$\chempot$'' proportional to the cosmological constant.
Moreover, the action \eqref{SN} is fully diffeomorphism invariant,
but the integrand $n(x) \propto \sqrt{-g(x)}$ is \emph{not},
as it is a scalar density.   

If we now consider this $n(x)$ to be a physical quantity
with $\chempot$ in \eqref{SN} interpreted as a chemical potential
(possibly related to an underlying spacetime crystal),
then $n(x)$ must be invariant under coordinate transformations
and this implies that the only allowed coordinate transformations
are those with unit Jacobian.
In that case, it is possible that $n$  also enters the matter 
potential $\epsilon$, as discussed
in Sec.~\ref{sec:Metric-determinant-as-dynamic-variable}.
It is precisely this last step which makes for the ``extension''   
mentioned in the title of the present paper.
(The possibility of adding extra $\sqrt{-g}$ terms
in the matter action was already noted 
on p. 220 of Ref.~\cite{Zee1983}, but was not pursued further.)

The example potential $\epsilon$ from \eqref{Expansion3} 
then shows that, for an appropriate \emph{range}
of $\chempot$ values, the equilibrium value of $n$ can nullify the total
gravitating vacuum energy density \eqref{Expansion4}.
In a cosmological context as discussed in
Secs.~\ref{sec:Cosmology-Basic-model}
and \ref{sec:Cosmology-Quantum-dissipative-effects}, the dynamics of
$n$ displays an attractor behavior towards Minkowski spacetime,
provided quantum-dissipative effects are taken into account. 
The cosmological cancellation of an initial vacuum energy density,
perhaps the most important result of this paper, is
illustrated by
Figs.~\ref{fig:num-sol-lambda0pt0001-deSbcs-gamma0-modODEs}
and \ref{fig:num-sol-lambda0pt0001-deSbcs-gamma2E11-modODEs}.

There are two ingredients for this cosmic reduction of an initial
vacuum energy density $\rho_\text{vac}$
(including a genuine cosmological constant $\Lambda$). 
First, the quantum-dissipative processes
give an energy transfer from the vacuum component
(with energy density $\rho_\text{vac}$ and
equation-of-state parameter $w_\text{vac}=-1$)
to a particle component (with $\rho_{M}$ and $w_{M}\geq 0$).
Second, the expansion of the Universe does not
affect the vacuum energy density [$\rho_\text{vac}(t)$ is constant]
but does reduce the matter energy density  [$\rho_{M}(t)$ drops 
with increasing cosmic scale factor $\widetilde{R}(t)$
from the Robertson--Walker  metric \eqref{eq:extRW-ds2}].
Such a two-step process has been considered 
before~\cite{KlinkhamerVolovik-MPLA-2016,Klinkhamer2022-preprint}. 
New, here, is that the vacuum variable is not a 
postulated quantity (such as a 4-form field strength  
or a 4D-brane density), 
but is provided by the already available spacetime metric, 
namely by its determinant.
\begin{acknowledgments}
The first part of this paper,
Secs.~\ref{sec:Crystal}--\ref{sec:Comparison-with-cond-mat},
was developed in collaboration with G.E. Volovik,
who also made valuable comments on the rest of the paper. 
The referee is thanked for constructive remarks. 
\end{acknowledgments}
\begin{appendix}
\newpage  
\section{ANALYTIC FRIEDMANN-TYPE SOLUTION FOR GENERAL $\mathbf{w_{M}}$} 
\label{app:Analytic-solution-wchi-general}

We present in this appendix
an exact solution of the ODEs \eqref{eq:dimensionless-ODEs}
for matter equation-of-state parameter $w_{M} > -1$
and $\chempot>0$  (similar results hold for $\chempot<0$).
As in the main text, 
we take the following \textit{Ansatz} functions for $\tau > 0$:%
\bsubeqs\label{eq:functions-analytic-sol-app}
\beqa
a(\tau) &=& \alpha\;\tau^{-2\,p} \,,
\\[2mm]
r(\tau) &=& \alpha^{-1/6}\;\widehat{r}\; \tau^{p/3} \,,
\\[2mm]
r_{\chi}(\tau) &=& \alpha^{-1}\;\widehat{\chi}\;\tau^{-m}\,,
\eeqa
\esubeqs
with positive parameters $\alpha$, $p$, $\widehat{r}$, $\widehat{\chi}$,
and $m$. 
The vanishing of $r_\text{vac}$ 
from \eqref{eq:dimensionless-ODE-rvac} gives  
\beq
\label{eq:rhat-analytic-sol-app}
\widehat{r}_\text{\,sol} =
\left[\frac{\,1}{2\,\zeta}\;\big(\chempot- \lambda\big)\right]^{1/3}\,,
\eeq
where, for a given value $\chempot>0$, the following condition holds
on the dimensionless cosmological constant $\lambda$:
\beq
\label{eq:lambda-condition-analytic-sol-app}
 \lambda < \chempot\,,
\eeq
so that $\lambda$ can also be negative
(see the last paragraph of this appendix for a related remark).

For the  \textit{Ansatz} functions \eqref{eq:functions-analytic-sol-app},
the dimensionless Ricci and Kretschmann curvature scalars read%
\bsubeqs\label{eq:R-K-analytic-sol-app}
\beqa
\label{eq:R-analytic-sol-app}
\mathcal{R}&=&
\frac{2}{3}\,
p\,\big(  5\,p -3 \big)\,\frac{1}{\alpha}\,{\tau}^{-2\,(1- p)}\,,
\\[2mm]
\label{eq:K-analytic-sol-app}
\mathcal{K}&=&
\frac{4}{27}\,p^{2}\,\big( 9 - 24\,p + 17\,p^{2}\big)\,
\frac{1}{\alpha^2}\,{\tau}^{-4\,(1- p)}\,.
\eeqa
\esubeqs
We look for an expanding ($p>0$) Friedmann-type universe
approaching Minkowski spacetime (different from a de-Sitter
spacetime at $p=1$).

With the \textit{Ansatz} functions \eqref{eq:functions-analytic-sol-app},
the three ODEs from \eqref{eq:dimensionless-ODEs}
reduce to the following expressions:
\bsubeqs\label{eq:ODES-functions-analytic-sol-app}
\beqa
\label{eq:ODES-rchieq-functions-analytic-sol-app}
0 &=&
\frac{1}{\alpha}\;
    \Big(  p\,\left(1 + w_{M}\right) - m \Big)\,
    \widehat{\chi}\,{\tau}^{-1 - m}  \,,
\\[2mm]
\label{eq:ODES-1stFeq-functions-analytic-sol-app}
0 &=&
\frac{p^{2}}{3\,{\tau}^{2}} - 8\,\pi\,\widehat{\chi} \,{\tau}^{-m - 2\,p}\,,
\\[2mm]
\label{eq:ODES-2ndFeq-functions-analytic-sol-app}
0 &=&
 \frac{p^{2}}{{\tau}^{2}} - \frac{2\,p}{3\,{\tau}^{2}} +
  8\,\pi\,w_{M}\,\widehat{\chi} \,{\tau}^{-m - 2\,p} \,.
\eeqa
\esubeqs
The exact solution has arbitrary $\alpha>0$ and
\bsubeqs\label{eq:constants-analytic-sol-app}
\beqa
p_\text{\,sol} &=& \frac{2}{3 + w_{M}}\,,
\\[2mm]
m_\text{\,sol} &=& \frac{2\,\left( 1 + w_{M} \right) }{3 + w_{M}}\,,
\\[2mm]
\widehat{\chi}_\text{\,sol} &=&
\frac{1}{6\,\pi \,{\left( 3 + w_{M} \right) }^{2}} \,,
\eeqa
\esubeqs
where $p_\text{\,sol}$ ranges over $(0,\,1)$
for $w_{M} \in (-1,\,+\infty)$. Phrased differently, the
analytic solution for fixed values of the model parameters
$\chempot$, $\lambda$, and $w_{M}$, has 
a one-dimensional modulus space 
$\mathbb{R}^{+}$ parametrized by $\alpha$.

The main points of the cosmology, with $w_{M}=1$ for example,
are as follows:  
\begin{itemize}  
  \item[(i)]
an expanding Friedmann-type universe with scale factor
$r \sim \tau^{1/6}$.\vspace*{-0.00mm}
  \item[(ii)]
a perfect-fluid energy density and
pressure $r_{\chi}(\tau)=p_{\chi}(\tau) \sim 1/\tau$.\vspace*{-0.00mm}
  \item[(iii)]
a cosmological constant cancelled
by $\sqrt{-g}=\widehat{r}_\text{\,sol}$
from \eqref{eq:rhat-analytic-sol-app},
provided condition \eqref{eq:lambda-condition-analytic-sol-app} holds,
so that $r_\text{vac}(\tau)=0$.  
\vspace*{-0.00mm}
  \item[(iv)]
the curvature scalars $\mathcal{R}(\tau) \sim 1/\tau$ and
$\mathcal{K}(\tau)\sim 1/\tau^{2}$.
\end{itemize}
The overall behavior of this $w_{M}=1$ cosmology is not very different
from that of Sec.~\ref{subsec:Analytic-solution-wchi-onethird},
which had a $w_{M}=1/3$ perfect fluid.

Let us end with a parenthetical remark 
expanding on the third item of the previous paragraph.  
It is, namely, possible to relax
condition \eqref{eq:lambda-condition-analytic-sol-app}
by changing
the $\epsilon$ $\textit{Ansatz}$ \eqref{eq:epsilon-Ansatz}.
An example, using dimensionless variables, is given by
$\epsilon=\lambda + n +1/n^2$,
which gives $r_\text{vac}=\lambda + 2\,n  - 1/n^2 -u$ from
\eqref{VacEnergy}.
In this $r_\text{vac}$ expression, \emph{any} value of $\lambda-u$
can be cancelled 
by an appropriate real value $n=\overline{n}>0$.

\section{ANALYTIC DE-SITTER-TYPE SOLUTION} 
\label{app:Analytic-solution-deS}

The ODEs \eqref{eq:dimensionless-ODEs}
have an analytic Friedmann-type solution, as discussed in
Sec.~\ref{subsec:Analytic-solution-wchi-onethird} 
and App.~\ref{app:Analytic-solution-wchi-general}.
But there is also an analytic de-Sitter-type solution,
which will be presented here.

As before, we assume the model parameters to obey the following conditions:
\bsubeqs\label{eq:conditions-analytic-deS-sol}
\beqa
\zeta &>& 0 \,,
\\[2mm] 
\chempot &>& 0 \,,
\\[2mm]
\lambda &<& \chempot\,,
\eeqa
\esubeqs
where the condition on the chemical potential $\chempot$ 
is only to simplify the discussion (what really matters is that
the combination $\chempot-\lambda$ is positive).
The \textit{Ansatz} functions for $\tau > 0$ are  
\bsubeqs\label{eq:functions-analytic-deS-sol}
\beqa
a(\tau) &=& \alpha\;\tau^{-2\,p} \,,
\\[2mm]
r(\tau) &=& \alpha^{-1/6}\;\widehat{r}\; \tau^{p/3} \,,
\\[2mm]
r_{\chi}(\tau) &=& 0\,,
\eeqa
\esubeqs
with positive parameters $\alpha$, $p$, and $\widehat{r}$.
The general de-Sitter-type solution (denoted ``deS-gen-sol'') then
has the following parameters:%
\bsubeqs\label{eq:constants-analytic-deS-gen-sol}
\beqa
p_\text{deS-gen-sol} &=& 1\,,
\\[2mm]
\alpha_\text{deS-gen-sol} &=& 1/(24\,\pi\,r_\text{vac-deS-gen-sol})\,,
\\[2mm]
r_\text{vac-deS-gen-sol} &=&
\lambda   + 2\,\zeta\,(\widehat{r}_\text{deS-gen-sol})^{3}  -\chempot >0\,,
\\[2mm]
\widehat{r}_\text{deS-gen-sol} &>& 0\,.
\eeqa
\esubeqs
The corresponding
dimensionless Ricci and Kretschmann curvature scalars read%
\bsubeqs\label{eq:R-K-analytic-deS-gen-sol}
\beqa
\label{eq:R-analytic-deS-gen-sol}
\mathcal{R}_\text{deS-gen-sol}&=&
\frac{4}{3}\,\frac{1}{\alpha_\text{deS-gen-sol}}\,,
\\[2mm]
\label{eq:K-analytic-deS-gen-sol}
\mathcal{K}_\text{deS-gen-sol}&=&
\frac{8}{27}\,\frac{1}{\alpha_\text{deS-gen-sol}^2}\,.
\eeqa
\esubeqs

The above solution has $r_\text{vac}$ (or, equivalently, $\widehat{r}$\,)
as a free parameter. For $\lambda>0$,  
a special solution (denoted  ``deS-spec-sol'') has 
vacuum energy density $r_\text{vac}=\lambda$ if
the following parameters are chosen:
\bsubeqs\label{eq:constants-analytic-deS-spec-sol} 
\beqa
p_\text{deS-spec-sol} &=& 1\,,
\\[2mm]
\alpha_\text{deS-spec-sol} &=& 
 \frac{1}{24\,\pi\,\lambda}\,,
\\[2mm]
\widehat{r}_\text{deS-spec-sol} &=& 
\sqrt[3]{\frac{\,\chempot}{2\,\zeta} }\,.
\eeqa
\esubeqs 
The dimensionless Ricci and Kretschmann curvature scalars 
are given by \eqref{eq:R-K-analytic-deS-gen-sol}
with $\alpha_\text{deS-gen-sol}$ replaced by $\alpha_\text{deS-spec-sol}$.

\section{READJUSTMENT AFTER A PHASE TRANSITION}  
\label{app:Readjustment-after-PhT}

The readjustment   
of the vacuum variable $q$ to a cosmological phase transition
has been discussed, in general terms, by Sec.~II-C
of Ref.~\cite{KlinkhamerVolovik2008a}.
We expect a similar behavior of the metric-determinant vacuum
variable $n$ from \eqref{eq:n-E-sqrtminusg}, 
especially as we have observed an attractor behavior
in the numerical results of Secs.~\ref{sec:Cosmology-Basic-model} 
and \ref{sec:Cosmology-Quantum-dissipative-effects}.
The present appendix aims to verify these expectations with
a simplified setup. From now on, we will use only the dimensionless
variables from Sec.~\ref{subsec:Cosmology-Basic-model-ODEs}.

Specifically, we model the phase transition 
by taking different $\lambda$ values
before and after a cosmic time $\tau_\text{PhT}$\,,  
\beq\label{eq:lambda-stepfunction}
\lambda(\tau)=
\begin{cases}
 \lambda_{1} \,,   &  \text{for}\;\; \tau<\tau_\text{PhT} \,,
 \\[2mm]
 \lambda_{2} \,,   &  \text{for}\;\; \tau>\tau_\text{PhT} \,.
\end{cases}
\eeq
Let us consider, for simplicity, the case $\lambda_{2}>\lambda_{1}>0$.
Then, we will solve the modified
ODEs \eqref{eq:dimensionless-modified-ODEs}
over the cosmic time interval $\tau \in [\tau_\text{bcs},\,\tau_\text{PhT})$
for $\tau_\text{bcs}<\tau_\text{PhT}$ 
and $\lambda=\lambda_{1}$
and over the interval
$\tau \in (\tau_\text{PhT} ,\,\tau_\text{max}]$
for $\tau_\text{PhT} <\tau_\text{max}$ 
and $\lambda=\lambda_{2}$.
At $ \tau=\tau_\text{PhT}$, 
we take the metric functions $\{ a(\tau),r(\tau),\dot{r}(\tau)\}$ 
to be continuous and
the $r_{\chi}$ value just above $\tau_\text{PhT}$ from
the first Friedman equation \eqref{eq:dimensionless-modified-ODEs-1stFeq}
with $\lambda=\lambda_{2} >\lambda_{1}$.
(Different matching conditions
are needed for the case $0<\lambda_{2} <\lambda_{1}$.)

\begin{figure}[b]   
\begin{center}
\hspace*{0mm}
\includegraphics[width=1\textwidth]{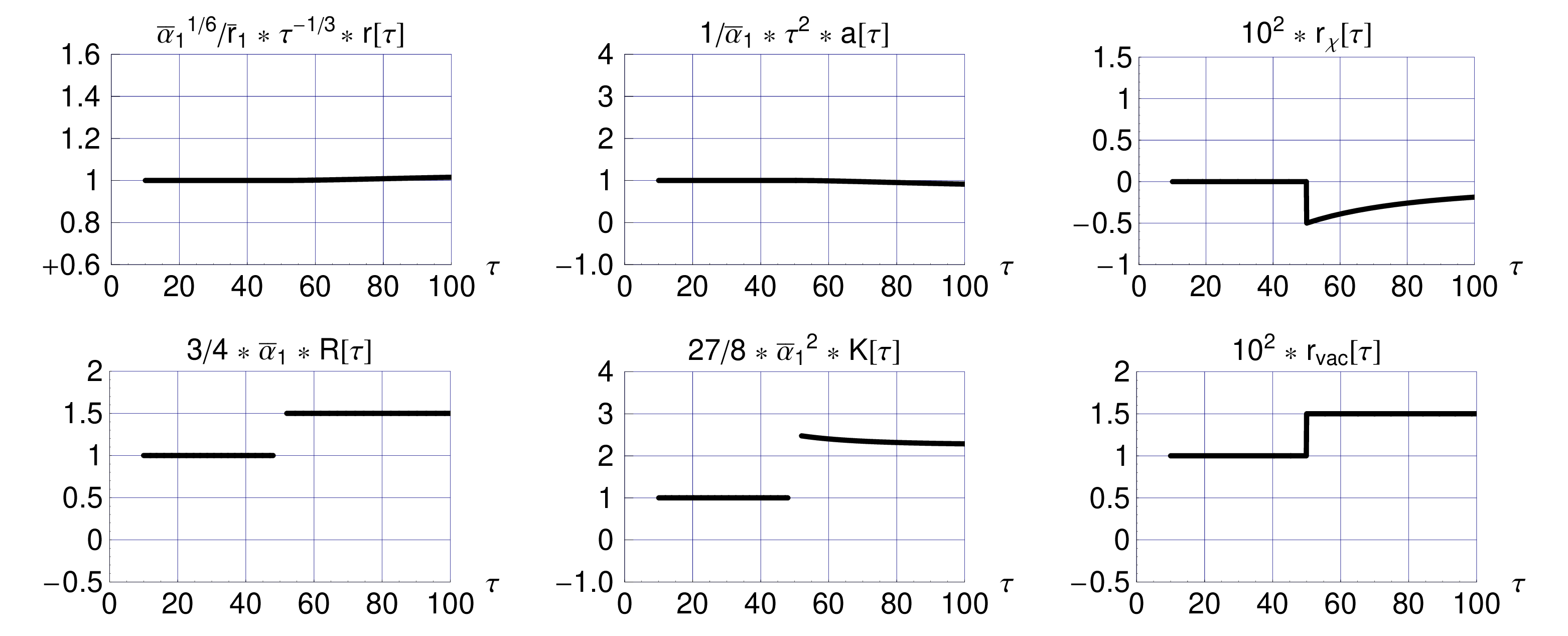}
\end{center}\vspace*{-0mm}
\caption{Numerical solution of the modified
ODEs \eqref{eq:dimensionless-modified-ODEs} with
the source term \eqref{eq:Gamma-tilde}
and the $\lambda$ stepfunction \eqref{eq:lambda-stepfunction}
modelling a cosmological phase transition 
at $\tau=\tau_\text{PhT}$, 
for parameters $w_{M}=1/3$, $\zeta=1$, $\chempot=3$, 
$\lambda_{1}=10^{-2}$, $\lambda_{2}=1.5 \times 10^{-2}$,
and $\gamma=0$ (quantum-dissipative effects turned off).
The initial boundary conditions at $\tau=\tau_\text{bcs}=10$
are taken from the analytic de-Sitter-type solution 
and are similar
to those of Fig.~\ref{fig:num-sol-lambda0pt0001-deSbcs-gamma0-modODEs}. 
The boundary conditions 
at $\tau=\tau_\text{PhT}{}^{+}=50^{+}$
take the numerical function values of $\{ a(\tau),r(\tau),\dot{r}(\tau)\}$ 
obtained just below $\tau_\text{PhT}$, 
while the $r_{\chi}(50^{+})$ value follows from
the first Friedman equation \eqref{eq:dimensionless-modified-ODEs-1stFeq}
with $\lambda=\lambda_{2}$.
}
\label{fig:num-sol-lambda-PhT-gamma0}
\end{figure}

\begin{figure}[p]
\begin{center}
\hspace*{0mm}
\includegraphics[width=1\textwidth]{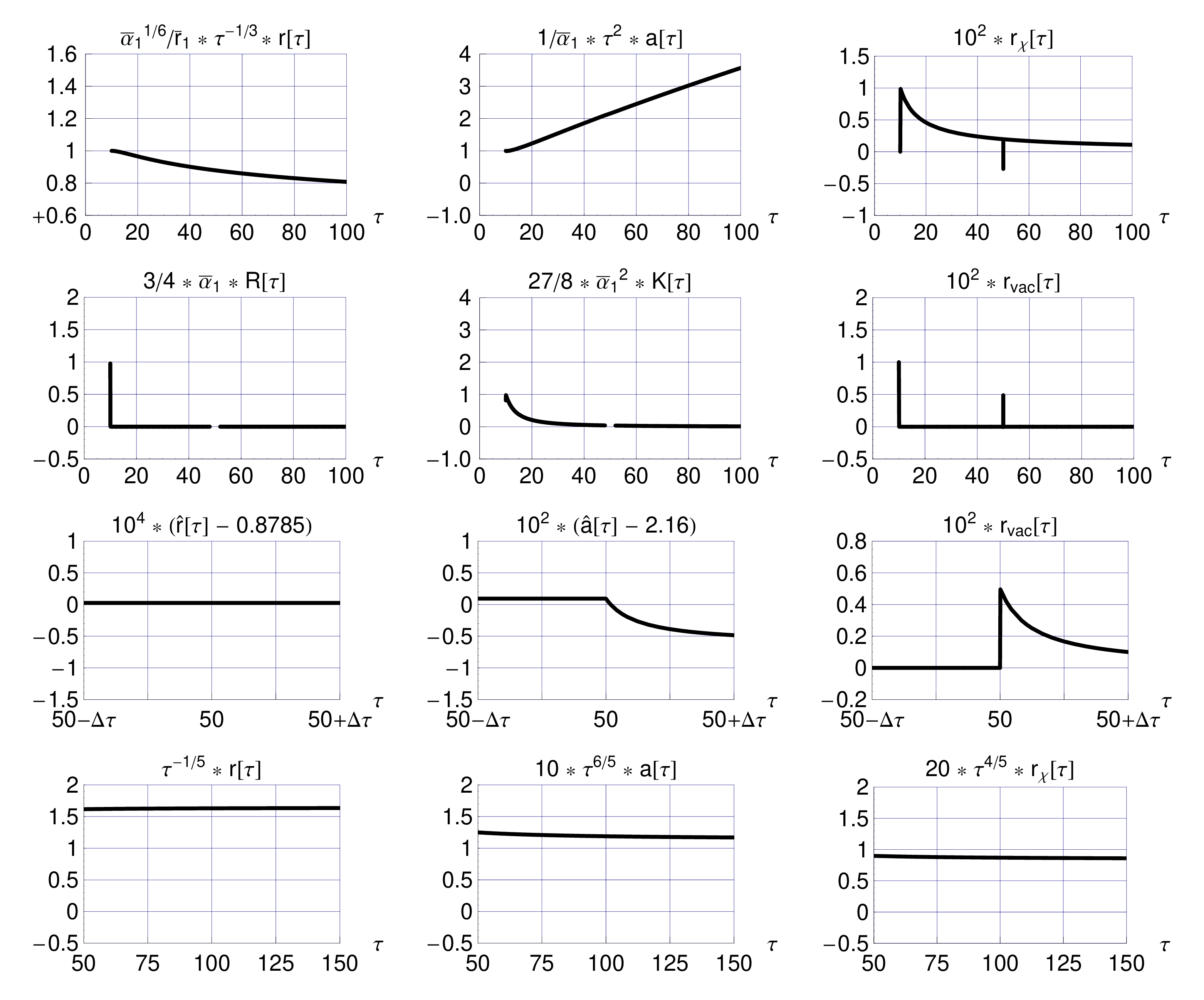}
\end{center}\vspace*{-0mm}
\caption{The boundary conditions at $\tau=\tau_\text{bcs}=10$
and the model parameters are the same as in
Fig.~\ref{fig:num-sol-lambda-PhT-gamma0},
but now with $\gamma=2 \times 10^{11}$ (quantum-dissipative effects turned on).
The vacuum energy density starts from the value
$r_\text{vac}(10)=1 \times 10^{-2}$,  
drops to $r_\text{vac}(50^{-})\approx 2 \times 10^{-10}$,
then jumps to $r_\text{vac}(50^{+}) \approx 0.5 \times 10^{-2}$,
and finally drops to $r_\text{vac}(100)\approx 3 \times 10^{-10}$.
The numerical results for $\mathcal{R}(\tau)$ and $\mathcal{K}(\tau)$ 
on the second row are not plotted at $\tau_\text{PhT}=50$, 
as the ``spikes'' there are primarily artifacts.
The third row shows the behavior of the two metric functions
and the vacuum energy density 
in a small interval around at $\tau_\text{PhT}=50$ with
$\Delta\tau=10^{-6}$
[the left and middle panels use the combinations 
$\widehat{r}(\tau)\equiv 
\overline{\alpha}_{1}{\,}^{1/6}/\overline{r}_{1}\,\tau^{-1/3}\,r(\tau)$ 
and 
$\widehat{a}(\tau)\equiv 1/\overline{\alpha}_{1}\,\tau^{2}\,a(\tau)$,
which are precisely the combinations from the top row].  
The fourth row shows the asymptotic behavior of the two metric functions
and the matter energy density. 
}
\label{fig:num-sol-lambda-PhT-gamma2E11}
\vspace*{0mm}
\end{figure}

Numerical results are shown in Fig.~\ref{fig:num-sol-lambda-PhT-gamma0} 
for the case without vacuum-matter energy exchange ($\gamma=0$) 
and in Fig.~\ref{fig:num-sol-lambda-PhT-gamma2E11} for the case 
with vacuum-matter energy exchange ($\gamma >0$).
The results from Fig.~\ref{fig:num-sol-lambda-PhT-gamma2E11}
demonstrate that, as expected, 
the added vacuum energy from a phase transition
can be rapidly cancelled by the metric-determinant vacuum variable,
provided quantum-dissipative effects are included.


\end{appendix}

\newpage


\begin{thebibliography}{99} 


\bibitem{Weinberg1989}
S. Weinberg,
``The cosmological constant problem,''
Rev. Mod. Phys.  \textbf{61}, 1 (1989).

\bibitem{Carroll2001}
S.M.~Carroll,
``The cosmological constant,''
Living Rev. Relativity \textbf{4}, 1 (2001),
arXiv:astro-ph/0004075.  

\bibitem{KlinkhamerVolovik2008a}
F.R. Klinkhamer and G.E. Volovik,
``Self-tuning vacuum variable and cosmological constant,''
Phys. Rev. D \textbf{77}, 085015 (2008), arXiv:0711.3170.

\bibitem{KlinkhamerVolovik2008b}
F.R.~Klinkhamer and G.E.~Volovik,
``Dynamic vacuum variable and equilibrium approach in cosmology,''
Phys. Rev. D \textbf{78}, 063528 (2008), arXiv:0806.2805.

\bibitem{KlinkhamerVolovik2019}
F.R. Klinkhamer and G.E. Volovik,
Tetrads and $q$-theory,
JETP Lett. \textbf{109},  364 (2019),
arXiv:1812.07046.

\bibitem{KlinkhamerVolovik2022-BBasTopQPT}
F.R.~Klinkhamer and G.E.~Volovik,
``Big bang as a topological quantum phase transition,''
Phys. Rev. D \textbf{105},  084066 (2022),
arXiv:2111.07962. 

\bibitem{NissinenVolovik2019}
J. Nissinen and G.E. Volovik,
Elasticity tetrads, mixed axial-gravitational anomalies,
and (3+1)-d quantum Hall effect,
Phys. Rev. Res. \textbf{1}, 023007 (2019),
arXiv:1812.03175.

\bibitem{Nissinen2020}
J.~Nissinen,
``Field theory of higher-order topological crystalline response,
generalized global symmetries and elasticity tetrads,''
arXiv:2009.14184. 

\bibitem{Einstein1919}  
A. Einstein,
``Spielen Gravitationsfelder im Aufbau 
der materiellen Elementar\-teilchen eine wesentliche Rolle?''
(Do gravitational fields play an essential role 
in the structure of the elementary particles of matter?),
Preussische Akademie der Wis\-sen\-schaf\-ten Berlin,
Sizungsberichte (Math. Phys.), \textbf{1919}, 349 
(1919); 
paper and translation available from                      
\verb"https://einsteinpapers.press.princeton.edu/vol7-doc/187" 
and 
\verb"https://einsteinpapers.press.princeton.edu/vol7-trans/101". 


\bibitem{vanderBij-etal1982}
 J.J. van der Bij, H. van Dam, and Y.J. Ng,
 ``The exchange of massless spin two particles,''
Physica (Amsterdam) \textbf{116} A, 307 (1982).

\bibitem{Zee1983}
A. Zee,
``Remarks on the cosmological constant problem,''
in \textit{High-Energy Physics: Proceedings of Orbis Scientiae 1983},  
edited by  S.L. Mintz and A. Perlmutter  
(Plenum Press, N.Y., 1985), pp. 211--230.

\bibitem{BuchmuellerDragon1988}
W. Buchm\"uller and N. Dragon,
``Einstein gravity from restricted coordinate invariance,''
Phys. Lett. B \textbf{207}, 292 (1988).

\bibitem{HenneauxTeitelboim1989}
M. Henneaux and C. Teitelboim,
``The cosmological constant and general covariance,''
Phys. Lett. B \textbf{222}, 195 (1989).

\bibitem{BensityGuendelman-etal2020}
D.~Bensity, E.I.~Guendelman, A.~Kaganovich, E.~Nissimov,
and S.~Pacheva,
``Non-canonical volume-form formulation of modified gravity theories and cosmology,''
Eur. Phys. J. Plus \textbf{136}, 46 (2021),
arXiv:2006.04063.  

\bibitem{AlvarezFaedo2007}
E.~Alvarez and A.F.~Faedo,
``Unimodular cosmology and the weight of energy,''
Phys. Rev. D \textbf{76}, 064013 (2007),
arXiv:hep-th/0702184.  


\bibitem{Mukhanov2005}  
V.~Mukhanov,       
\emph{Physical Foundations of Cosmology}
(Cambridge University Press, Cambridge, England, 2005).

\bibitem{Bamba-etal2016}
K.~Bamba, S.D.~Odintsov, and E.N.~Saridakis,
``Inflationary cosmology in unimodular $F(T)$ gravity,''
Mod. Phys. Lett. A \textbf{32}, 1750114 (2017),
arXiv:1605.02461.  

\bibitem{Klinkhamer2017}
F.R. Klinkhamer,
``A generalization of unimodular gravity with vacuum-matter energy exchange,''
Int. J. Mod. Phys. D \textbf{26},  1750006 (2016),
arXiv:1604.03065. 


\bibitem{KlinkhamerSavelainenVolovik2016}
F.R.~Klinkhamer, M.~Savelainen, and G.E.~Volovik,
``Relaxation of vacuum energy in $q$-theory,''
J. Exp. Theor. Phys. \textbf{125},  268 (2017),
arXiv:1601.04676.  

\bibitem{ZeldovichStarobinsky1977}
Ya.B. Zel'dovich and A.A. Starobinsky,
``Rate of particle production in gravitational fields,''
JETP Lett. \textbf{26}, 252 (1977); paper available 
from \verb"http://jetpletters.ru/ps/1379/article_20902.pdf".   

\bibitem{KlinkhamerVolovik-MPLA-2016}
F.R.~Klinkhamer and G.E.~Volovik,
``Dynamic cancellation of a cosmological constant
and approach to the Minkowski vacuum,''
Mod. Phys. Lett. A \textbf{31}, 1650160 (2016),
arXiv:1601.00601.  

\bibitem{Klinkhamer2022-preprint}
F.R.~Klinkhamer,
``Q-field from a 4D-brane:
Cosmological constant cancellation and Min\-kow\-ski attractor,''
to appear in LHEP,  
arXiv:2207.03453. 






\end{thebibliography}
\end{document}